# Placement-driven reversal of preferred spin alignment in monolayer hBN vacancy pairs

Daniel Hashemi*

Department of Physics, Optical Engineering and Nanoengineering, Rose-Hulman Institute of Technology, Terre Haute, Indiana 47803, United States

*E-mail: hashemi@rose-hulman.edu

## ABSTRACT

Negatively charged boron vacancies, $V_{\mathrm{B}}^{-}$, are optically addressable spin centres and quantum sensors in hexagonal boron nitride (hBN). Close vacancy pairs interact electronically as well as magnetically. Using density-functional theory at total charge $q = -2$, we show that in monolayer hBN the preferred collinear spin alignment of such a pair depends on atomic placement, and its strength does not follow the separation. The closest pair, 4.3 Å apart, favours parallel alignment by 27.6 meV within the semilocal approximation. Moving one vacancy by a single lattice vector reverses the preference, and a more distant 7.5 Å pair favours antiparallel alignment by 50.7 meV, more strongly than the nearer reversed pair. The reference pairs differ in connectivity: a compact three-nitrogen contact versus a single N–B–N–B–N chain. In larger supercells, numerical controls change their splittings by less than 1 meV; a hybrid functional retains the signs while substantially changing the magnitudes.

Atomically thin hosts place spin defects close to the materials they probe. In hexagonal boron nitride (hBN), the negatively charged boron vacancy, $V_{\mathrm{B}}^{-}$, combines a spin-triplet ground state with optical initialization and readout at room temperature,[1–4] and layers containing these centres have enabled magnetic imaging of neighbouring materials.[5,6] Irradiation and ion implantation create the defects,[7,8] and dense ensembles show strongly interacting spin dynamics, described in terms of magnetic dipolar coupling between spatially distributed spins.[8] A different regime becomes accessible when two vacancies approach within a few lattice constants: their electronic states can then interact as well as their magnetic moments. A natural first guess is that the preferred spin alignment of such a pair, and its strength, are set by the distance between the vacancies. In this Letter we show that, in monolayer hBN, separation is not a reliable guide: one lattice step reverses the alignment, and its strength varies non-monotonically with distance.

Figure 1 summarizes the central result. The closest pair we consider, at lattice placement (1,1) and 4.3 Å, favours parallel (P) alignment. Displacing one vacancy by a single primitive lattice vector, to (2,1), reverses the preference to antiparallel (AP), and the more distant (3,0) pair, at 7.5 Å, prefers AP more strongly than the nearer (2,1) pair. Moving the partner changes how the vacancies are connected through the lattice, not only how far apart they are. The two reference pairs differ in connectivity: at (1,1) the vacancies meet through a compact three-nitrogen contact, whereas at (3,0) a single N–B–N–B–N chain joins them. The opposite signs of (1,1) and (3,0), and of the neighbouring placements (1,1) and (2,1), persist under all numerical controls performed and under the HSE06 hybrid functional, which nevertheless changes the magnitudes substantially. Throughout, we therefore distinguish whether a number is reproducible within a chosen approximation (numerical precision) from how it depends on that approximation (accuracy with respect to the functional).

When two vacancies share a monolayer, the nitrogen dangling-bond states that carry their unpaired electrons[2] can mix directly or through the intervening lattice. Previous calculations established magnetism in neutral-vacancy hBN sheets[9] and spacing-dependent magnetic ordering in periodic vacancy arrays,[10,11] and adjacent boron vacancies can reconstruct into a fused complex.[12,13] Spin–optical coupling has been studied in donor–acceptor pairs,[14,15] work on diamond showed that defect-pair exchange depends on orbital symmetry and direction as well as distance,[16] and interlayer boron-vacancy pairs show reported sub-meV alignment splittings at the same nominal total charge.[17] Here, two non-adjacent boron vacancies share one monolayer, and only their relative lattice placement is varied at fixed total charge.

We use spin-polarized density-functional theory in the Perdew–Burke–Ernzerhof (PBE) approximation[18] with the projector augmented-wave method[19,20] as implemented in VASP,[21] compare with the screened hybrid functional HSE06[22,23] at fixed geometry, and test the angular dependence with constrained noncollinear calculations.[24] Section S1 of the Supporting Information gives computational details.

We label the partner-vacancy position by $\mathbf{\Delta} = n\mathbf{a}_1 + m\mathbf{a}_2$, where the primitive vectors enclose 60° and the lattice constant is $a = 2.505$ Å (Figure 1a), so that the ideal separation is $d = a\sqrt{n^2 + nm + m^2}$. Each placement is calculated separately, with one pair in an $N \times N$ supercell; the seven separations range from 4.339 to 10.020 Å (Table 1; Figure S1). We exclude the adjacent (1,0) configuration, whose vacancies share a nitrogen neighbour and can reconstruct.[12,13] Every supercell contains two electrons more than the neutral defective cell, giving $q = -2$ with a uniform compensating background; this fixes the total charge, not a separate charge on each vacancy.

For each placement, we obtain P and AP collinear Kohn–Sham solutions at fixed total spin imbalances $N_\uparrow - N_\downarrow = 4$ and 0, that is, spin moments of $4\mu_\mathrm{B}$ and zero. Both branches use the same cell, total charge, numerical settings and nuclear coordinates, taken from the P-relaxed geometry, and we define

$$\Delta E = E_\mathrm{AP} - E_\mathrm{P}, \quad (1)$$

A positive $\Delta E$ therefore favours P. These self-consistent determinants are not, in general, eigenstates of total spin, so $\Delta E$ does not directly give a quantum multiplet spectrum. We assign a preferred alignment when $|\Delta E| \geq 0.5$ meV and call smaller differences weak; this reporting threshold is comparable to the 0.51 meV sampling change at (1,1) and is not a universal error bound.

Within PBE, the (1,1) pair favours P alignment by 27.64 meV (Figure 1b; Table 1). Moving the partner to (2,1), one primitive lattice vector away, reverses the preference to AP by 7.96 meV. The (3,0) pair favours AP more strongly, by 50.71 meV, while (3,1) and (4,0) favour AP by 2.20 and 5.24 meV in the 12×12 supercell. All vacancies occupy the boron sublattice, yet the signs differ; this does not conflict with Lieb's theorem,[25] whose half-filled, bipartite Hubbard-model assumptions do not describe this charged, multi-orbital system. The remaining placements are the weak (2,2) pair, with $\Delta E = +0.21$ meV, and (2,0), with $\Delta E = -377.0$ meV.

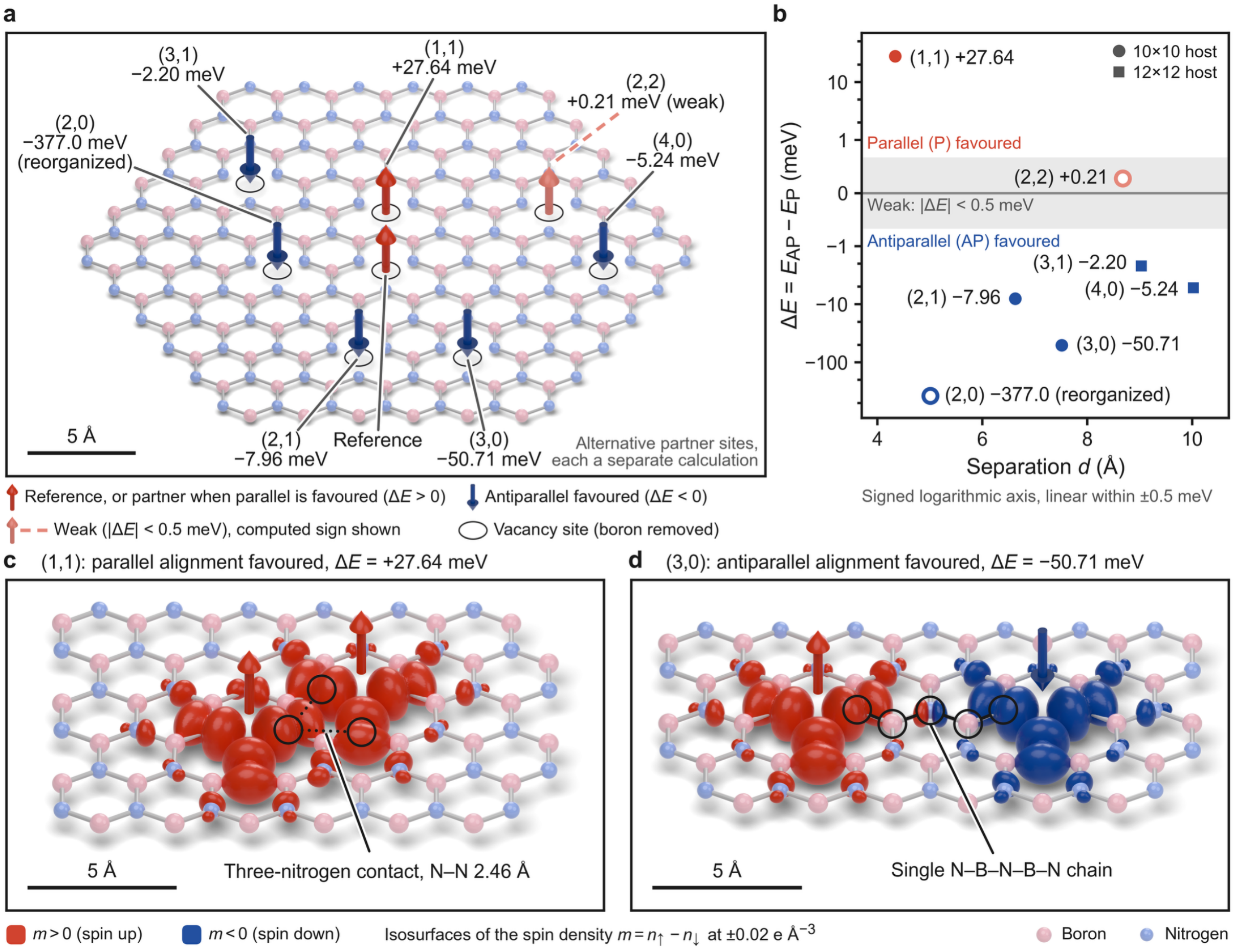

**Figure 1.** Atomic placement selects the preferred collinear alignment. (a) Partner sites relative to one reference vacancy; each arrow is a separate two-vacancy calculation at $n\mathbf{a}_1 + m\mathbf{a}_2$. Red upward and blue downward arrows denote a lower parallel (P) or antiparallel (AP) branch; pale red and a dashed connector mark the weak (2,2) result, $|\Delta E| < 0.5$ meV. (1,1) and (2,1) differ by one lattice vector despite being drawn at symmetry-equivalent positions. (b) $\Delta E = E_{\mathrm{AP}} - E_{\mathrm{P}}$ versus ideal separation (PBE, $q = -2$, P-relaxed nuclei) in 10×10 (circles) and 12×12 (squares) supercells; open markers identify (2,2) and the electronically reorganized (2,0). The grey band marks the weak interval, not an uncertainty; no distance dependence is fitted. (c,d) Spin-density isosurfaces ($\pm 0.02$ e Å$^{-3}$) for (1,1) P and (3,0) AP, enclosing 84% and 82% of $\int |m|\, dV$; black rings mark the three-nitrogen contact and the five-atom chain.

The (2,0) pair shows why a branch splitting should not automatically be called an exchange constant. Its facing dangling-bond nitrogens approach to 2.39 Å across one boron atom, its projected vacancy moments shrink by 16% from P to AP, and calculations without a fixed total spin moment find a third, $2.000\mu_{\mathrm{B}}$ solution lying between the two branches (Section S1.5). AP remains lowest among the solutions examined, also in hybrid-functional tests, but the energy difference includes changes of electronic occupation and moment magnitude. We therefore keep (2,0) in the alignment map without converting its splitting into a rigid-spin exchange parameter.

The magnitude does not follow the separation (Figure S2). Within the 10×10 calculations, (3,0) at 7.5 Å exceeds (2,1) at 6.6 Å by more than a factor of six; within the 12×12 set, (4,0) at 10.0 Å exceeds (3,1) at 9.0 Å, so no monotonic distance-decay law describes these placements. Because no two placements share a separation, and the only assigned parallel preference lies at the closest placement, the signs alone would also fit a single distance threshold and the map does not separate orientation from distance; what the magnitudes establish is that the strength of the preference cannot be inferred from the separation. Even the smallest assigned preference, 2.20 meV, is more than two orders of magnitude above the isolated-centre zero-field splitting of approximately 14 μeV,[1] and the point-dipole prefactor for two electron-spin moments 4.34 Å apart is only about 2.6 μeV. The resolved pairs therefore lie well beyond a dipolar-only description.

**Table 1.** Preferred collinear alignment across the seven placements. PBE values of $\Delta E = E_{\mathrm{AP}} - E_{\mathrm{P}}$ compare branches at identical P-relaxed coordinates, cell, total charge $q = -2$ and numerical settings; positive values favour parallel alignment. Differences below 0.5 meV are labelled weak, with their calculated sign retained. $d$ refers to the removed boron sites in the ideal lattice.

| **Placement $(n, m)$** | **$d$ (Å)** | **Super cell** | **$\Delta E$ (meV)** | **Preferred alignment** | **HSE06 sign** | **Notes** |
|---|---|---|---|---|---|---|
| (1,1) | 4.339 | 10×10 | +27.64 | Parallel | Kept (8×8) | Constrained angular energies depart from the bilinear form (8×8 supercell) |
| (2,0) | 5.010 | 10×10 | −377.0 | Antiparallel | Kept (6×6) | Includes an electronic reorganization; not converted into an exchange constant |
| (2,1) | 6.628 | 10×10 | −7.96 | Antiparallel | Kept (8×8) | 8×8 supercell: −7.89 meV |
| (3,0) | 7.515 | 10×10 | −50.71 | Antiparallel | Kept (10×10) | 12×12 supercell: −51.32 meV; one interior angle, consistent with the bilinear form |
| (2,2) | 8.678 | 10×10 | +0.21 | Weak | Not computed | Computed sign: parallel; one supercell only |
| (3,1) | 9.032 | 12×12 | −2.20 | Antiparallel | Not computed | One supercell only |
| (4,0) | 10.020 | 12×12 | −5.24 | Antiparallel | Not computed | One supercell only |

The 8×8 and 10×10 calculations use the tight settings (520 eV plane-wave cut-off, $10^{-8}$ eV electronic convergence) and the 12×12 calculations the standard settings (450 eV, $10^{-7}$ eV). The HSE06 column states whether the PBE sign persists at matched coordinates in the indicated supercell. Entries describe collinear branch ordering, not quantum multiplets.

We next test the precision and accuracy of these signs for the reference placements (1,1) and (3,0) (Figure 2). Periodic images matter: the (1,1) splitting decreases from +105.32 meV in the 48-atom 5×5 cell to +30.76, +27.60 and +27.64 meV in the 6×6, 8×8 and 10×10 cells, so the smallest cell overestimates the 10×10 value by a factor of 3.8, whereas the final size step changes it by only 0.04 meV. The 5×5 cell, which also shows strong sampling sensitivity and an additional low-moment solution (Sections S1.2 and S2.2), therefore represents a dense periodic array rather than an isolated pair; we do not assign the same error to other hBN calculations.[15,17,26] In the larger reference cells, the numerical controls performed change the two principal splittings by at most 0.61 meV (Table S3). Coverage is more limited elsewhere: (2,1) was checked in two supercells, (2,0) in four, and (2,2), (3,1) and (4,0) in one each.

Except at (2,0), each vacancy's projected moment in the reference supercell lies between 1.28 and $1.40\mu_{\mathrm{B}}$ in both branches, within 5% of the isolated-vacancy value; this operational criterion, chosen after inspecting the moments, is consistent with two nominal $V_{\mathrm{B}}^{-}$ centres but does not establish their individual charge states (Table S2). Allowing the AP branch to relax gives relaxed-branch differences of +26.11 and −53.30 meV for (1,1) and (3,0) (Section S3.1); the opposite preferences therefore survive branch relaxation and are not produced by comparing different geometries.

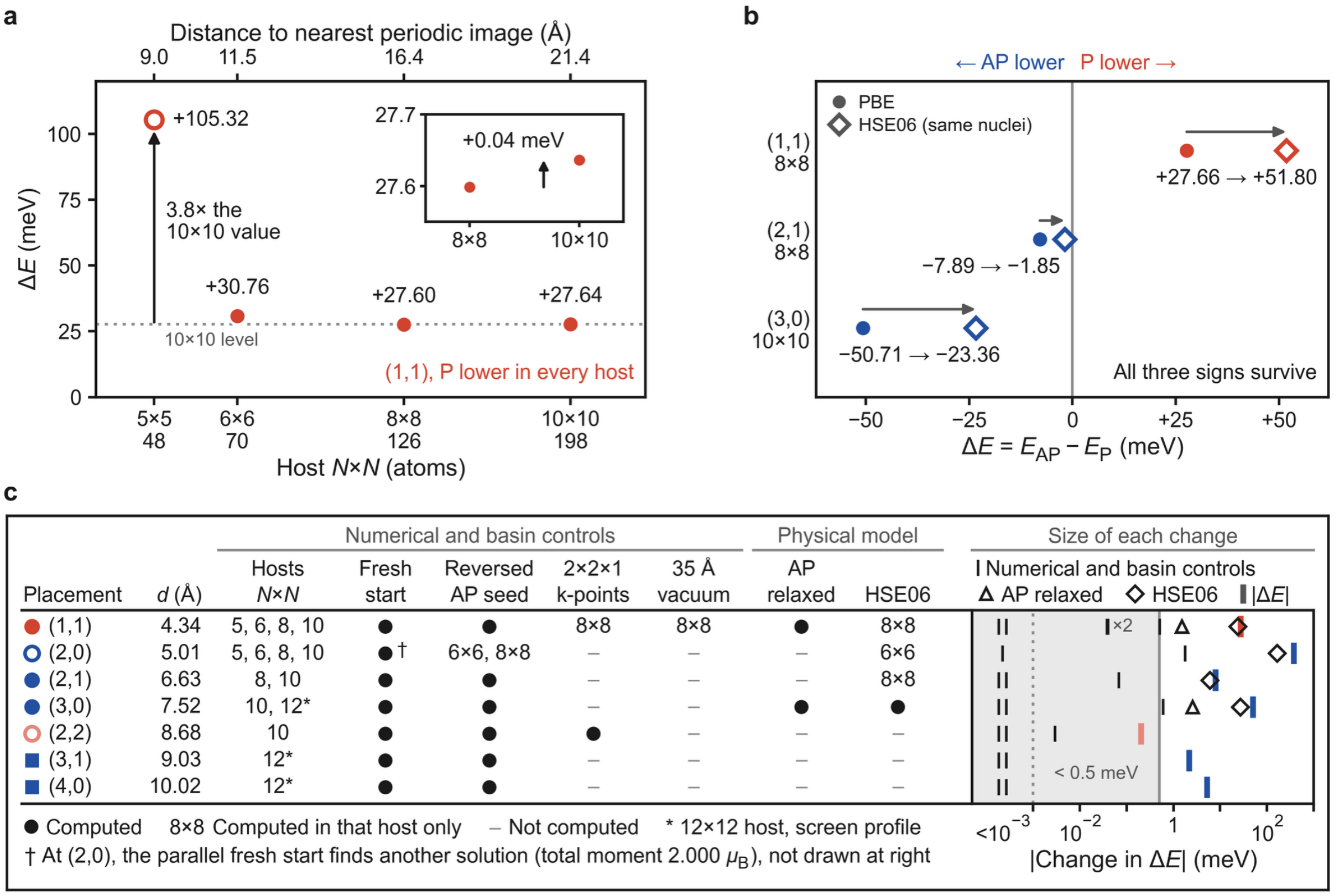

**Figure 2.** Numerical precision and functional dependence. (a) PBE splitting of (1,1) versus supercell size (upper axis: nearest periodic-image distance); the 5×5 value is 3.8 times the 10×10 result, whereas the final size increment changes $\Delta E$ by 0.04 meV. The open marker flags an additional AP solution in the smallest cell. (b) Matched-geometry PBE and HSE06 splittings: all signs persist, although HSE06 shifts each value towards P and reverses the magnitude ranking of the reference pairs. Supercells differ between the reference placements; arrows indicate changes of approximation, not uncertainty bars. (c) Coverage and magnitude of the controls (Table S3): dots mark performed tests, not outcomes; bars show $|\Delta E|$; the grey region denotes values below 0.5 meV; changes below $10^{-3}$ meV lie left of the dotted line; ×2 marks nearly coincident values.

HSE06 produces a much larger change but preserves every tested sign. At the same PBE coordinates, $\Delta E$ changes from +27.66 meV (the PBE state used to initialize HSE06; Section S1.2) to +51.80 meV for (1,1) in 8×8, from −7.89 to −1.85 meV for (2,1) in 8×8, and from −50.71 to −23.36 meV for (3,0) in 10×10 (Figure 2b); (2,0) also retains AP ordering (Section S1.5). The neighbouring placements (1,1) and (2,1) therefore keep opposite signs, and the more distant (3,0) pair still prefers AP more strongly than the nearer (2,1) pair, albeit in a larger supercell, whereas the magnitude ranking of (1,1) and (3,0) reverses. Agreement of signs is the central robustness result; the functional shifts are neither error bars nor evidence that either magnitude is exact. No hybrid calculations were performed for (2,2), (3,1) or (4,0), whose weaker PBE preferences remain predictions to be tested at that level.

Why might one lattice step reverse the sign? A simple orbital picture organizes the analysis. The isolated vacancy has a filled $a_1'$ level and a half-filled $e'$ doublet derived from nitrogen dangling bonds,[2] so a pair supplies four frontier $e'$ levels per spin channel. In P these are all occupied in the majority channel and empty in the minority channel, and mixing two filled levels leaves their sum unchanged in an ideal two-level model with fixed onsite energies. In AP, each spin channel contains occupied states on one vacancy and empty states on the other; their mixing lowers the occupied levels, the familiar route to an AP-favouring kinetic contribution.[27] This is a qualitative guide, not a decomposition of the self-consistent energy, because onsite potentials, other orbitals and double-counting terms also respond.

The two reference pairs realize this mixing through different connections (Figures 1c,d and 3). At (3,0), the unique shortest path is a four-bond N–B–N–B–N chain, and the nearest dangling-bond nitrogens of the two vacancies remain 4.89 Å apart. At (1,1), one nitrogen of vacancy A lies 2.46 Å from two nitrogens of vacancy B, forming a compact three-nitrogen contact joined by two N–B–N bridges. Both show same-spin level mixing in P, with effective two-level couplings of approximately 0.23 eV at (3,0) and 0.15–0.17 eV at (1,1) (Section S4.3).

We probe these connections by displacing selected atoms through $\delta = 0.02$ Å and recomputing both branches. Mode definitions, acceptance checks and descriptor predictions were fixed before these calculations, although the first-order energy slopes were already known from the undisplaced forces, so the tests concern electronic responses rather than blind energy predictions (Section S6). For (3,0), flattening the chain changes $\Delta E$ from −57.28 through −50.71 to −44.63 meV at $-\delta$, zero and $+\delta$, a half-span response of −12.47% in its magnitude, and both same-spin couplings decrease, by 7.3% and 5.8%, within the predicted range. The chain bonds shorten, yet the AP preference weakens, so bond length alone does not control the interaction. A structurally matched off-path control changes the energy by only 0.054% on the same scale. A chain-end control shifts the relative onsite levels by 0.33 and 0.58 of the chain response but changes $\Delta E$ by only 6% of it, so at first order the onsite shift alone does not account for the energy response. Neither control reproduces the occupied-channel coupling response (Section S6.4).

These results support an occupied–empty hybridization contribution through the chain without isolating it as the complete mechanism. The predicted decrease of the AP–P Kohn–Sham gap contrast is not observed (it increases by 5.7%); the chain moment also decreases by 5–6%, so a contribution quadratic in that moment could vary on a comparable scale; the chain-end control reproduces the empty-channel coupling response to 0.45; and the occupied–empty admixture and the squared two-level coupling rank (1,1) and (3,0) in opposite orders (Section S4.4). HSE06, which opens the same-spin gap at (3,0) and weakens its AP preference, is compatible with a gap-sensitive contribution but does not by itself single one out (Section S6.7).

The closest pair has a different, spatially concentrated signature. At fixed nuclei, the three contact-nitrogen cells carry $0.140\mu_B$ less spin magnitude in AP than in P, 62% of the whole-cell contrast, while the intervening hollow shows no corresponding deficit; an occupied contact-localized state loses most of its spin splitting, from 0.379 to 0.045 eV (Figure 3c). The facing-atom contrast is 2.7 times smaller at (3,0) and at least 14 times smaller at (2,1) and (2,2), and the signature persists after AP relaxation, in 8×8 and with HSE06 (Section S6.7).

This localized response makes the contact a natural focus for explaining the P preference, but it does not identify a separate contact energy: a filled-state splitting reports a local spin-dependent potential, and reversing one of two overlapping spin densities can reduce their combined magnitude without changing either profile. Across four (1,1) distortion modes, two of which lower and two raise $\Delta E$, neither contact descriptor nor the occupied–empty proxy keeps a consistent single-variable relation to the energy (Figure 3d; Section S6.5). The calculations thus establish a reproducible, contact-specific electronic signature whose energetic contribution they do not isolate.

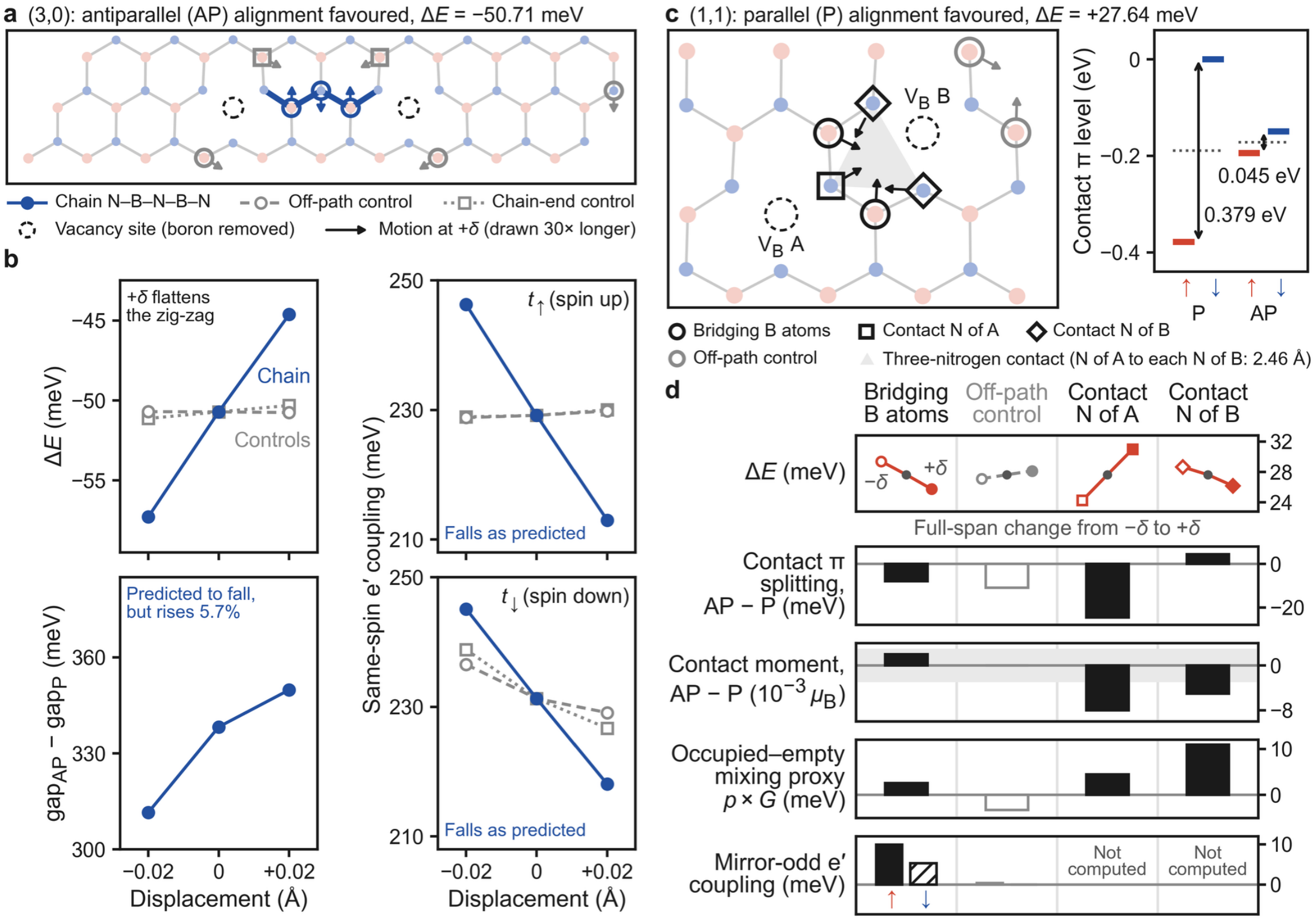

**Figure 3.** Local distortions distinguish chain and contact responses. (a) Displacement modes for (3,0): the chain mode moves the three interior atoms and flattens the zig-zag; off-path and chain-end modes are controls. Arrows show the +0.02 Å displacement, enlarged 30-fold. (b) Energy splitting, P-branch same-spin couplings $t_\uparrow$ and $t_\downarrow$, and gap contrast $\mathrm{gap_{AP}} - \mathrm{gap_P}$. Each control changes the splitting by less than 1 meV over the full span; the gap contrast increases, contrary to its prediction. (c) Contact modes for (1,1) and the occupied contact-localized levels, whose spin splitting decreases from 0.379 eV in P to 0.045 eV in AP. (d) Full-span responses of the four (1,1) modes: $\Delta E$, contact splitting and moment contrasts, and mixing descriptors. Open and filled endpoints denote negative and positive displacement; grey bands mark numerical thresholds. Collinear PBE at $q = -2$.

Converting a collinear splitting into an exchange constant requires a spin model. For two rigid spin-1 centres with $H = -J\mathbf{S}_1 \cdot \mathbf{S}_2$, collinear product states give $\Delta E = 2J$; the lowest singlet-to-triplet excitation for $J < 0$ is then $|J|$, not $|\Delta E|$,[28,29] and the classical angular energy is linear in $\cos\theta$. We test this form with noncollinear PBE without spin–orbit coupling, constraining the directions of the six nitrogen moments bordering the vacancies[24] and evaluating the penalty-removed energy at the angle actually reached. The endpoints reproduce the collinear splittings computed in the same supercell within 0.13 meV for (1,1) and 0.07 meV for (3,0).

For (1,1) in 8×8, the constrained energies lie systematically above the bilinear interpolation (Figure 4a). At penalty strength $\lambda = 20$ eV $\mu_B^{-1}$, the residuals are 2.2, 4.4, 4.3 and 3.3 meV near 45°, 90°, 112.5° and 135°; the near-112.5° point was added after a held-out test had failed. That test used a three-term form, $c_0 + c_1\cos\theta + c_2\cos^2\theta$, fixed before the held-out calculations, which misses the near-135° point by 1.05 meV. Vacancy B's projected moment varies by 8.9%, exceeding the 5% rigidity criterion, whereas A varies by 2.2%. Because 79–93% of the penalty at interior angles enforces alignment within each vacancy's nitrogen triad, the energies may include a cost of suppressed internal relaxation. The result, further limited by the finite penalty strength, Γ-point sampling and an incomplete larger-cell check, is a departure from the bilinear form for these atom-wise-constrained PBE states, not a constraint-independent quantum spin Hamiltonian (Section S5). At (3,0) in 10×10, the single interior angle lies within 0.6 meV of the interpolation and the projected moments vary by approximately 2% (Figure 4c), consistent with, but not a validation of, a bilinear description; because the two placements use different supercells, we interpret their angular results separately.

We therefore retain $\Delta E$ as the primary observable and use $J \equiv \Delta E/2$ only as a conditional endpoint parameter. Under the isotropic two-spin-1 mapping, whose validity is not established here, (3,0) would have a singlet ground state with a triplet approximately 25 meV above it in PBE or 12 meV in HSE06, whereas (1,1) would have a quintet ground state, although its angular results do not justify a quantitative bilinear multiplet ladder. With $k_B T \approx 26$ meV at 300 K, several multiplets would be thermally occupied (Section S5.5).

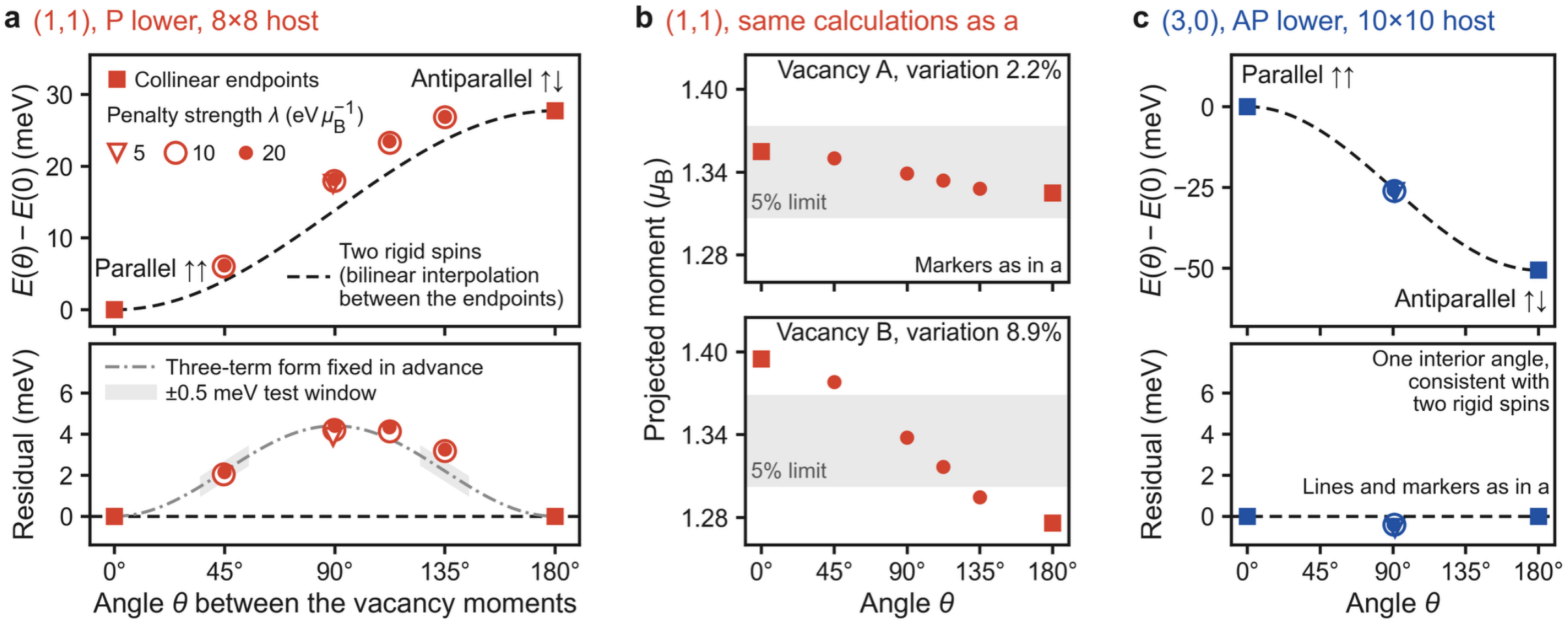

**Figure 4.** The closest pair departs from a bilinear angular-energy curve under atom-wise constraints. (a) Penalty-removed noncollinear PBE energies of (1,1) in 8×8 versus the achieved angle between vacancy moments. The dashed curve, $1/2\,\Delta E(1 - \cos\theta)$, joins the endpoints without fitting (noncollinear splitting 27.725 meV). The grey curve, the three-term form fixed from the endpoints and the near-90° point, misses the near-135° energy by 1.05 meV; grey windows mark its ±0.5 meV criterion. (b) Projected vacancy moments along the path vary by 2.2% (A) and 8.9% (B), against a predefined 5% rigidity criterion; endpoint squares are from the $\lambda = 5$ calculations, where the penalty vanishes. (c) (3,0) in 10×10: the single interior residual, approximately −0.6 meV at $\lambda = 5$ and −0.4 meV at $\lambda = 10$ and 20, is compatible with the bilinear interpolation without validating it.

The (3,0) preference and same-spin coupling thus respond together to distortion of the connecting chain, whereas the closest pair concentrates its spin-magnitude contrast at a compact contact. These signatures do not uniquely partition the total energy into microscopic exchange terms, and the failed or intervention-dependent descriptor predictions sharpen the next question: which constrained electronic variable can isolate the contact or chain contribution to the energy?

For spin-defect engineering, the pairs with assigned preferences lie in an interaction regime beyond independent centres or dipolar coupling, in which the sign depends on lattice placement. Models of close-pair subsets in dense, strongly interacting ensembles[8] should therefore retain lattice-resolved connectivity rather than assign interactions from a monotonic distance law. For predictive work, both the supercell and the functional must be stated: periodic images strongly distort small-cell values, and sub-meV numerical precision does not imply functional accuracy, particularly for the weaker preferences untested with HSE06.

Experimentally, the contrast motivates searches for deliberately created[7,8] or individually located close pairs, followed by low-temperature magnetic or spin-resonance characterization; within a valid two-spin-1 description, the conditional 12–25 meV singlet–triplet scale of (3,0) would make its magnetic response thermally activated. Translating this scale into a resonance frequency or optical signal requires pair-specific anisotropy, selection rules and optical-transition calculations, and neither a singlet assignment nor a contrast in optically detected magnetic resonance follows from the collinear branch ordering alone. Abundance is a separate question: at the reported parts-per-million charged-vacancy concentrations,[8] randomly distributed subnanometre pairs would be uncommon, and implantation can also create vacancies in other charge states, so the fixed-$q = -2$ calculation represents one possible electronic configuration rather than an ensemble's composition. Formation energies, charge-transition levels[30,31] and migration or reconstruction barriers would connect this energy landscape to experimental populations.

In summary, a change of lattice placement alone, without changing the host or the defect species, can reverse the preferred collinear spin alignment of two boron vacancies in monolayer hBN; the reversal persists under HSE06, and the strength of the preference does not follow the separation. The chain and contact geometries identified here provide reference configurations for lattice-resolved models of coupled spin defects in two dimensions and concrete targets for deliberately created pairs.

## ASSOCIATED CONTENT

### Supporting Information

Computational methods (structural models, numerical settings, branch preparation, total charge and hybrid-functional settings); numerical controls, projected moments and splitting versus separation; relaxation and hybrid-functional checks; electronic structure at fixed nuclei; constrained noncollinear calculations; deformation tests; exploratory bond-resolved analysis and reduced models; analysis chronology (PDF)

## AUTHOR INFORMATION

### Corresponding Author

Daniel Hashemi – Department of Physics, Optical Engineering and Nanoengineering, Rose-Hulman Institute of Technology, Terre Haute, Indiana 47803, United States; Email: hashemi@rose-hulman.edu

### Notes

The author declares no competing financial interest.

## Data Availability

The source data underlying the figures and tables, the raw data and the reproducibility record are deposited under DOI 10.5281/zenodo.22994971; the source data identify the plotted or tabulated values, units and originating calculations. The deposition is not a complete archive of wavefunctions and charge densities, and the licensed VASP PAW potentials are not redistributed. VASP 6.6.1 is licensed software; LOBSTER 5.1.1 is used for the bond analysis (Section S7) and Wannier90 3.1.0 for the exploratory localized-orbital construction (Section S8.1). Input-generation, parsing, descriptor, evaluation and figure scripts are available from the author on reasonable request; the figure workflow uses Python, matplotlib, scikit-image and Blender. The computational procedures and numerical settings are documented in the Supporting Information and the reproducibility record.

## ACKNOWLEDGMENTS

Computational resources included Anvil at Purdue University through ACCESS allocation MAT260089. ACCESS is supported by U.S. National Science Foundation grants 2138259, 2138286, 2138307, 2137603 and 2138296.[32,33] This work also used the Oak Ridge Leadership Computing Facility at Oak Ridge National Laboratory, supported by the U.S. Department of Energy Office of Science under Contract DE-AC05-00OR22725.

## For Table of Contents Only

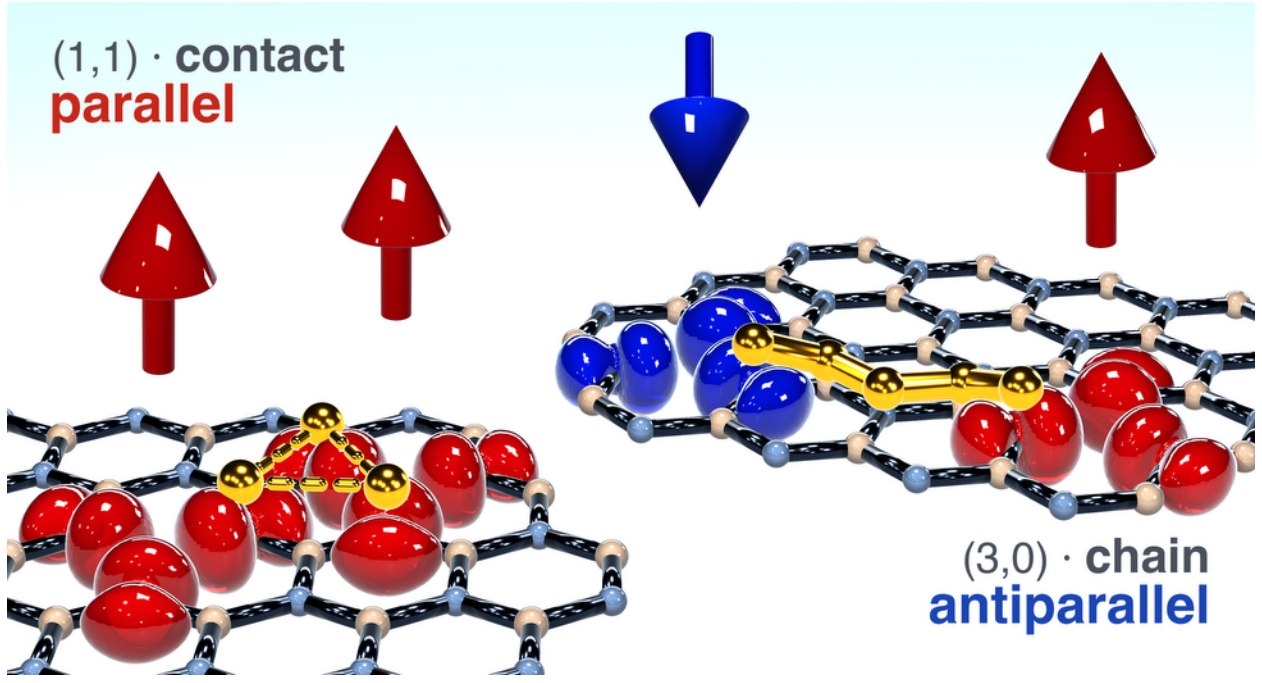

# Supporting Information

## Placement-driven reversal of preferred spin alignment in monolayer hBN vacancy pairs

Daniel Hashemi*

Department of Physics, Optical Engineering and Nanoengineering, Rose-Hulman Institute of Technology, Terre Haute, Indiana 47803, United States

*E-mail: hashemi@rose-hulman.edu

**Contents**

Number of pages: 33; Figures S1–S7; Tables S1–S15

**Conventions.** $\Delta E = E_{\mathrm{AP}} - E_{\mathrm{P}}$ compares antiparallel (AP) and parallel (P) collinear Kohn–Sham branches at identical nuclei (P-relaxed unless stated), cell, total charge $q = -2$ and settings. Positive $\Delta E$ favours P; the branches are not spin-pure multiplets. Differences below 0.5 meV are labelled weak, with their computed sign kept, as for (2,2) at +0.21 meV; this is a reporting convention, not an uncertainty. Atom labels such as N99 index the stated placement's own 10×10 structure and do not transfer between structures. A value below its numerical floor is not determined at that sensitivity; it is not necessarily zero and does not show that a contribution is absent.

# S1. Computational methods: models, settings, branch preparation and total charge

## S1.1. Placements, supercells and numerical settings

Monolayer hBN is represented by hexagonal $N \times N$ supercells with $N = 5$, 6, 8, 10 and 12, lattice constant $a = 2.505$ Å and out-of-plane repeat 25 Å. Removing two boron atoms leaves 48–286 atoms per pair cell; the supercells used for each placement are listed in Section S2.2. Their separation vector is $\mathbf{\Delta} = n\mathbf{a}_1 + m\mathbf{a}_2$, with 60° between primitive vectors, giving $d = a\sqrt{n^2 + nm + m^2}$. The adjacent (1,0) pair is excluded because it shares a nitrogen neighbour and can reconstruct. The closest retained configuration is (1,1), at $\sqrt{3}a = 4.339$ Å.

The seven configurations (Figure S1) sample the 0° and 30° directions and two intermediate directions, with separations up to 10.020 Å. Every calculation contains one pair and every vacancy occupies the boron sublattice. The map therefore tests relative placement within one sublattice, not a change of defect species or sublattice. Reference supercells and nearest-image distances are given in Section S2.1. P-branch structures are relaxed to a maximum force below 0.03 eV Å$^{-1}$ (0.015–0.030 eV Å$^{-1}$ in the individual relaxations; Section S3.1). For (1,1) and (3,0), the AP branch is relaxed separately, and each branch is evaluated at both relaxed geometries (Section S3.1).

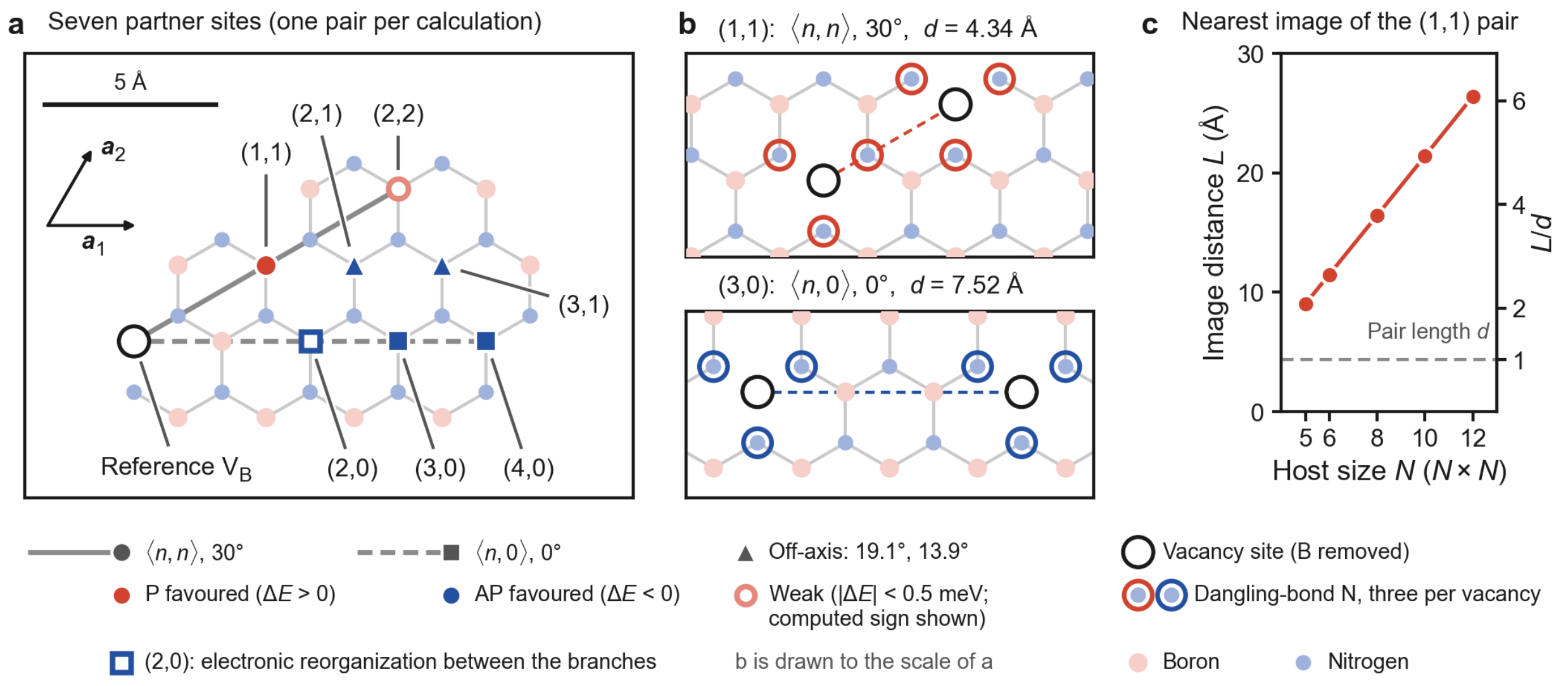

**Figure S1. Lattice placements and periodic-image distances. (a)** The seven partner sites relative to a fixed vacancy: circles mark the 30° family, squares the 0° family and triangles the off-axis sites at 19.1° and 13.9°; red denotes P and blue AP preference, the pale open (2,2) symbol a weak splitting and the open (2,0) symbol electronic reorganization. **(b)** The (1,1) and (3,0) structures with their nitrogen neighbours highlighted. **(c)** Nearest periodic-image distance $L$ versus supercell size for (1,1), with $L/d$ on the right axis; the image lies approximately 2.1 pair lengths away in 5×5 and 4.9 in 10×10.

**Electronic structure.** Spin-polarized PBE calculations[1] use the projector augmented-wave (PAW) method[2,3] in VASP 6.6.1,[4] with PAW potentials for the $2s^2 2p^1$ (B) and $2s^2 2p^3$ (N) valence configurations. No symmetry is imposed, and the Brillouin zone is sampled at Γ unless a 2×2×1 mesh is stated. We use two sets of numerical settings. The tight settings, with a 520 eV plane-wave cut-off, 0.02 eV Gaussian smearing and $10^{-8}$ eV electronic convergence, serve the principal fixed-geometry calculations through 10×10. The standard settings use a 450 eV cut-off and 0.03 eV smearing, with $10^{-5}$ eV electronic convergence in relaxations and $10^{-7}$ eV in the 12×12 fixed-geometry calculations; the 12×12 cells use the standard settings because of their size. We report energies extrapolated to zero smearing. Hybrid calculations use 25% or 32% exact exchange (Section S3.2).

**Precision of the settings.** At (3,0), changing supercell and settings together, from 10×10 with the tight settings to 12×12 with the standard settings, changes the splitting from −50.71 to −51.32 meV, that is, by 0.61 meV; this comparison does not separate the change of supercell from the change of settings. Except in the 5×5 (1,1) reference calculations, the PAW projection operators are evaluated in real space; in the smallest cells, evaluating them in reciprocal space instead changes $\Delta E$ by at most 0.0001 meV (Section S2.2). Selected small-cell relaxations and fixed-geometry calculations, and the six 198-atom states recomputed for the bond analysis (Section S7), were performed with a CPU build of VASP, and most other calculations with a GPU build of the same version; for identical small-cell inputs the two builds agree within $10^{-6}$ eV.

## S1.2. Branch preparation and repeated initializations

Different initial conditions can converge to different self-consistent solutions. We therefore start from initial moments of $\pm 2/3\mu_B$ on each of the three nitrogens surrounding a vacancy, a total of $2\mu_B$ per vacancy, with equal signs on the two vacancies in P and opposite signs in AP. Both branches are computed at fixed total spin moment, with the spin-up and spin-down electron numbers differing by 4 in P and 0 in AP. This is a magnetization constraint, not a projection onto a total-spin eigenstate (a spin-pure state). Calculations without the constraint, which we call unconstrained, are initialized from superposed atomic densities; continuing converged orbitals with the constraint removed is a different test, performed only for (2,0) in 6×6 (Section S1.5).

The Kohn–Sham equations are solved with the blocked Davidson algorithm for P calculations up to 126 atoms and with all-band conjugate-gradient minimization for larger P calculations and for all AP branches; the settings of individual calculations are given in the reproducibility record deposited with the raw data (Section S8.3). Reproducibility is tested in three ways: for AP, the direct sign assignment is compared with its site-reversed counterpart, in which the initial moments of the two vacancies are interchanged; unequal initial moments are placed on the two vacancies at (1,1) in 8×8; and unconstrained calculations test whether the same branch is recovered.

We keep the lowest converged solution for each branch; repeated recovery shows reproducibility over the initializations examined, not proof of a global minimum. In the 5×5 (1,1) cell, an unconstrained calculation from AP initial moments converges to a second solution 14.58 meV (≈14.6 meV) below the constrained AP solution, with total moment $0.82\mu_B$; we did not recover it in 6×6 or larger cells. This is the additional AP solution marked by the open symbol in Figure 2a of the main text. With the nearest image only 9.03 Å away and a strongly sampling-dependent splitting, the 5×5 cell is a dense periodic-pair calculation rather than the isolated-pair limit.

For (1,1) in 8×8, nine AP calculations differ in initial moments, in whether the spin constraint is applied and in the minimization algorithm. Eight of them, including both unconstrained calculations, agree within $2 \times 10^{-8}$ eV; an independent site-reversed calculation is 0.062 meV lower and supplies the accepted AP energy, a spread far below 0.5 meV. The resulting splitting, +27.60 meV, differs from the +27.66 meV of the directly initialized AP state used to initialize the matched HSE06 calculation; we keep the two references separate.

Two calculations that miss the electronic convergence criterion are retained only as diagnostics: the 6×6 P calculation with reciprocal-space projection, which reached $10^{-7}$ rather than $10^{-8}$ eV, and the unconverged 5×5 P calculation entering the approximate sampling-sensitivity range (both in Section S2.2). The reported alignment map uses converged vertical branch differences at P-relaxed nuclei.

## S1.3. Energy definition and the two-spin mapping

Energies are extrapolated to zero smearing and compared at identical nuclei, cell, charge and settings; $\Delta E$ is a collinear branch difference in the broken-symmetry framework.[5,6]

For an assumed pair of rigid spin-1 centres, $H = -J\mathbf{S}_1 \cdot \mathbf{S}_2$ gives $\Delta E = 2J$ between collinear product states. Its multiplet energies are $E(S_{\text{tot}}) = -J[S_{\text{tot}}(S_{\text{tot}} + 1) - 4]/2$. Positive $J$ gives a quintet ground multiplet. Negative $J$ gives a singlet, with the first triplet at $|J| = |\Delta E|/2$. This factor of two is essential when comparing a branch splitting with a model excitation energy.

We use $J \equiv \Delta E/2$ only as an endpoint definition under this model. Similar projected vacancy moments are an operational check, not sufficient to validate spin-1 centres. At (2,0) electronic reorganization and moment changes make the mapping inappropriate; at (1,1) the constrained angular curve departs from the bilinear interpolation, whereas at (3,0) one interior point is compatible with it. Neither result supplies a spin-pure spectrum.

## S1.4. Total charge and the charge-state scan

We add two electrons to the neutral defective supercell, giving total charge $q = -2$, compensated by a uniform background, and impose no local charge on either vacancy. We apply no charged-defect electrostatic correction, because the present observables compare branches at the same total charge rather than formation energies or charge-transition levels. This does not guarantee cancellation of all finite-size effects; their measured dependence is assessed through the supercell tests of Section S2. A fixed-geometry charge scan at (1,1) provides a separate comparison of local moments and occupations without assigning thermodynamic charge stability.

In this scan, at fixed (1,1) 8×8 geometry (Table S1), $q = -1$ and 0 give total moments of 5.000 and $6.000\mu_B$, whereas $q = -3$ gives a fractional moment of $4.515\mu_B$ and partially occupied frontier levels. A neutral isolated vacancy converges to a $1.0\mu_B$ branch under the corresponding procedure. These totals are therefore not simple sums of independently assigned local spins.

At $q = -2$, the vacancy projections of both reference branches lie within 5% of the isolated charged-vacancy projection. This supports two nominal charged centres as a working description without establishing their individual charge states. We did not calculate formation energies or charge-transition levels, which require a charged-slab electrostatic treatment.[7,8]

**Table S1. Fixed-geometry charge scan for (1,1) in 8×8.** The $q = 0, -1, -3$ states are initialized with parallel moments and computed without a spin-imbalance constraint; the two $q = -2$ rows are the reference branches. No electrostatic finite-size correction is applied, so these energies give no formation energies or transition levels.

| total charge q | electrons in cell | energy at zero smearing (eV) | total moment (μB) | per-vacancy moment (μB) | self-consistency iterations |
|---|---|---|---|---|---|
| 0 | 506 | -1092.810283 | 6.000 | +1.750/+1.760 | 41 |
| −1 | 507 | -1097.038829 | 5.000 | +1.628/+1.640 | 29 |
| −2 (reference, parallel branch) | 508 | -1099.707837 | 4.000 | +1.356/+1.395 | 25 |
| −2 (reference, antiparallel branch) | 508 | -1099.680177 | 0.000 | +1.325/-1.275 | 27 |
| −3 | 509 | -1100.387625 | 4.515 | +1.245/+1.345 | 153 |

**Structural persistence.** Adjacent vacancies can form a fused complex,[9,10] but its reported neutral binding energy does not establish the stability of the separated $q = -2$ pairs considered here. Predicting their persistence after preparation would require relative minima, charged-slab effects and migration or reconstruction barriers, which we have not computed; irradiation may also create configurations outside thermal equilibrium.

## S1.5. Three self-consistent solutions at (2,0)

The (2,0) pair has the largest PBE splitting, with AP lower by 377.0 meV in 10×10. Its facing nitrogens are 2.389 Å apart across one boron, compared with the two 2.457 Å contacts of (1,1). The projected vacancy moments change from 1.371 to $1.151\mu_B$ between P and AP (Table S2), that is, from $1.37\mu_B$ to $1.15\mu_B$ per vacancy when rounded, a 16% reduction. Unconstrained calculations converge to a third solution with total moment $2.000\mu_B$ and projections near $0.667\mu_B$ per vacancy.

In 10×10 PBE, this third solution lies 186.9 meV below P and approximately 190.1 meV above AP (187 and 190 meV when rounded). The highest occupied majority dangling-bond level of P, an antibonding combination concentrated on N99 and N109, is empty in the $2\mu_B$ solution, while the bonding combination is filled in the opposite spin channel. The facing nitrogens then carry moments near $-0.044\mu_B$, unlike the other four neighbouring nitrogens, so the comparison involves a change of occupation as well as orientation.

With a fixed spin imbalance, a single Fermi level does not mark the occupation boundary in each spin channel; we identify filled states from the occupations and do not compare levels referred to different calculations' Fermi levels as physical shifts.

In 6×6, calculations continued from the converged orbitals with the spin constraint removed retain the PBE $4\mu_B$ state and all three HSE06 solutions. In HSE06 the AP, P and asymmetric $2\mu_B$ states lie at 0, 238.79 and 359.99 meV relative to AP, so the third state lies below P in PBE but above it in HSE06. AP is lowest among the tested solutions in both functionals, a branch ordering rather than an exhaustive search for the global ground state. Between PBE and HSE06 the splitting of these 6×6 branches changes from −403.86 to −238.79 meV (Section S3.2; Table S3).

The PBE 6×6 P solution holds approximately 0.123 electron in a vacuum-like orbital, so its integer total moment alone does not establish localized spin-1 centres, and the HSE06 vacancy projections change by approximately 7% between P and AP. We therefore report the ordering and its functional dependence without deriving an exchange constant for this placement.

## S2. Numerical controls

For the two reference placements, no tested numerical change in the larger cells exceeds 0.61 meV; coverage is not uniform across placements, and we keep numerical precision distinct from functional accuracy.

### S2.1. Reference supercells and projected moments

The reference calculations use 10×10 supercells (198 atoms, 520 eV cut-off) or 12×12 supercells (286 atoms, 450 eV cut-off), with Γ-point sampling and a 25.0 Å out-of-plane repeat. Nearest periodic images lie 21.40 Å away for (1,1), 20.04 Å for (2,0), 18.91 Å for (2,1), 17.54 Å for (3,0) and 18.06 Å for (2,2) in 10×10, and 21.40 Å for (3,1), 20.04 Å for (4,0) and 22.54 Å for (3,0) in 12×12. (2,2), (3,1) and (4,0), each computed in a single supercell, have nearest-image distances of 18.1–21.4 Å; image distance alone is not evidence of convergence.

Projected moments indicate whether a branch comparison retains similar local electronic character. Our endpoint criterion requires every PAW-projected vacancy moment to lie within 5% of the isolated value and unconstrained calculations to reproduce the constrained splitting and total P moment; it is a diagnostic we chose after inspecting the moments, not a necessary-and-sufficient definition of a two-spin-1 system.

**Table S2. Vacancy projections and endpoint consistency.** Moments are PAW projections summed over the three nitrogens of vacancy A or B; the isolated reference is 1.338–1.340 μB. Changes compare magnitudes, and the deviation is the largest $\left|\ |m|-1.339\right|/1.339$. The criterion requires deviations below 5%, agreement of the unconstrained $\Delta E$ within 0.01 meV and a P total moment within 0.01 μB of the constrained value. It checks the collinear endpoints only, not an angular or multiplet model.

| (n,m) | supercell, cut-off | $m_P$, A / B (μB) | $m_{AP}$, A / B (μB) | change P→AP, A / B (%) | largest deviation from isolated (%) | unconstrained $\|\Delta E_{unc} - \Delta E\|$ (meV) | unconstrained $M_P$ (μB) | criterion |
|---|---|---|---|---|---|---|---|---|
| (1,1) | 10×10, 520 eV | 1.356 / 1.396 | +1.325 / −1.276 | 2.3 / 8.6 | 4.7 | < 0.001 | 4.000 | met |
| (2,0) | 10×10, 520 eV | 1.371 / 1.371 | −1.151 / +1.151 | 16.0 / 16.0 | 14.0 | 186.879 | 2.000 | not met (a) |
| (2,1) | 10×10, 520 eV | 1.339 / 1.340 | +1.334 / −1.330 | 0.4 / 0.7 | 0.7 | < 0.001 | 4.000 | met |
| (3,0) | 10×10, 520 eV | 1.340 / 1.340 | −1.313 / +1.313 | 2.0 / 2.0 | 1.9 | < 0.001 | 4.000 | met |
| (2,2) | 10×10, 520 eV | 1.337 / 1.337 | +1.337 / −1.337 | 0.0 / 0.0 | 0.1 | < 0.001 | 4.000 | met |
| (3,1) | 12×12, 450 eV | 1.327 / 1.326 | +1.325 / −1.325 | 0.2 / 0.1 | 1.0 | < 0.001 | 4.000 | met |
| (4,0) | 12×12, 450 eV | 1.326 / 1.326 | +1.323 / −1.323 | 0.2 / 0.2 | 1.2 | < 0.001 | 4.000 | met |

(a) The unconstrained (2,0) P calculation converges to the distinct $2.000\mu_B$ state, 186.9 meV below the constrained P solution, and fails the endpoint criterion.

The other five placements change their vacancy projections by at most approximately 2% between branches. At (1,1), vacancy B changes by 8.6%, yet both endpoint values lie within 5% of the isolated projection, one above and one below it; (1,1) therefore passes the endpoint comparison but fails the separate angular-rigidity test, which measures a different property.

## S2.2. Magnitude and coverage of the controls

Table S3 separates numerical and solution-basin controls from physical-model changes and marks untested cases, so that a small change at one placement is not generalized to the whole map.

**Table S3. Magnitude and coverage of the controls.** Entries are absolute changes in $\Delta E$, in meV. The final supercell step is 8×8 to 10×10 where available, or 10×10 (520 eV) to 12×12 (450 eV) for (3,0). Unconstrained calculations and site-reversed AP initializations probe reproducibility. Sampling and vacuum checks use matched references, except that the 35 Å calculation includes a separate relaxation. AP relaxation and HSE06 are physical-model comparisons. "Not computed" does not denote a zero change.

| (n,m) | reference supercell / cut-off | supercells computed | last supercell step | unconstrained calculation | site-reversed AP initialization | 2×2×1 k-points | 35 Å vacuum | AP relaxed ($\lambda_{opp}$) | PBE → HSE06 |
|---|---|---|---|---|---|---|---|---|---|
| (1,1) | 10×10 / 520 eV | 5, 6, 8, 10 | 0.038 | < 0.001 | < 0.001 | 0.507 (8×8) | 0.040 (8×8) | 1.528 | 24.135 (8×8) |
| (2,0) | 10×10 / 520 eV | 5, 6, 8, 10 | 1.784 | P: 186.879 (converges to the 2.000 μB solution); AP: < 0.001 | < 0.001 (6×6, 8×8) | not computed | not computed | not computed | 165.071 (6×6) |
| (2,1) | 10×10 / 520 eV | 8, 10 | 0.068 | < 0.001 | < 0.001 | not computed | not computed | not computed | 6.035 (8×8) |
| (3,0) | 10×10 / 520 eV | 10, 12 | 0.606 | < 0.001 | < 0.001 | not computed | not computed | 2.590 | 27.349 (10×10) |
| (2,2) | 10×10 / 520 eV | 10 | not computed | < 0.001 | < 0.001 | 0.003 (10×10) | not computed | not computed | not computed |
| (3,1) | 12×12 / 450 eV | 12 | not computed | < 0.001 | < 0.001 | not computed | not computed | not computed | not computed |
| (4,0) | 12×12 / 450 eV | 12 | not computed | < 0.001 | < 0.001 | not computed | not computed | not computed | not computed |

For (2,0) in 10×10, the AP calculation with the direct initial moments did not converge; the site-reversed solution we use is supported by converged calculations with both assignments in 6×6 and 8×8.

**Projector evaluation.** Evaluating the PAW projectors in real rather than reciprocal space shifts the total energies by approximately −14.09 meV in 5×5 and −20.84 meV in 6×6, equally in both branches, leaving the splittings of +105.320 and +30.761 meV unchanged within 0.0001 meV. This supports real-space evaluation in the larger cells without asserting identical cancellation for every structure. The reciprocal-space 6×6 P diagnostic reached an energy residual near $10^{-7}$ eV rather than its $10^{-8}$ eV target, much smaller than the measured offset; we retain the calculation only with this qualification.

**Unconstrained calculations.** At the same geometry and charge, they reproduce the accepted energies within $5 \times 10^{-7}$ eV and the total moments to the reported precision. The exception is P at (2,0), which reaches the 2 μB solution of Section S1.5, lower than the constrained P state by 223.03, 183.88 and 186.88 meV in 6×6, 8×8 and 10×10. Recovery shows that a state exists without the constraint, not that no lower solution exists.

**Size and sampling at (1,1).** The splitting is +105.32, +30.76, +27.60 and +27.64 meV in 5×5, 6×6, 8×8 and 10×10; the last two agree within 0.04 meV.

Increasing the sampling to 2×2×1 changes the 8×8 splitting by 0.51 meV, compared with an approximate 30–40 meV range in 5×5 that includes an unconverged P calculation and is not a converged sampling correction. A 35 Å out-of-plane repeat with a separate relaxation changes the accepted 8×8 result by 0.04 meV.

The P projections decrease from approximately 1.49 μB per vacancy in 5×5 to 1.36–1.40 μB in 8×8 and 10×10, and the AP projections there are approximately 1.325 and 1.276 μB in magnitude, close to but not identical with the isolated-vacancy projection.

### S2.3. Splitting versus separation

The placement map (Figure S2) contains two clear departures from monotonic decay: the 7.5 Å splitting exceeds the 6.6 Å value within the 10×10 set, and the 10.0 Å value exceeds the 9.0 Å value within the 12×12 set. No pair of placements has the same separation. The data therefore establish non-monotonic placement dependence, not a controlled comparison of orientation at fixed distance.

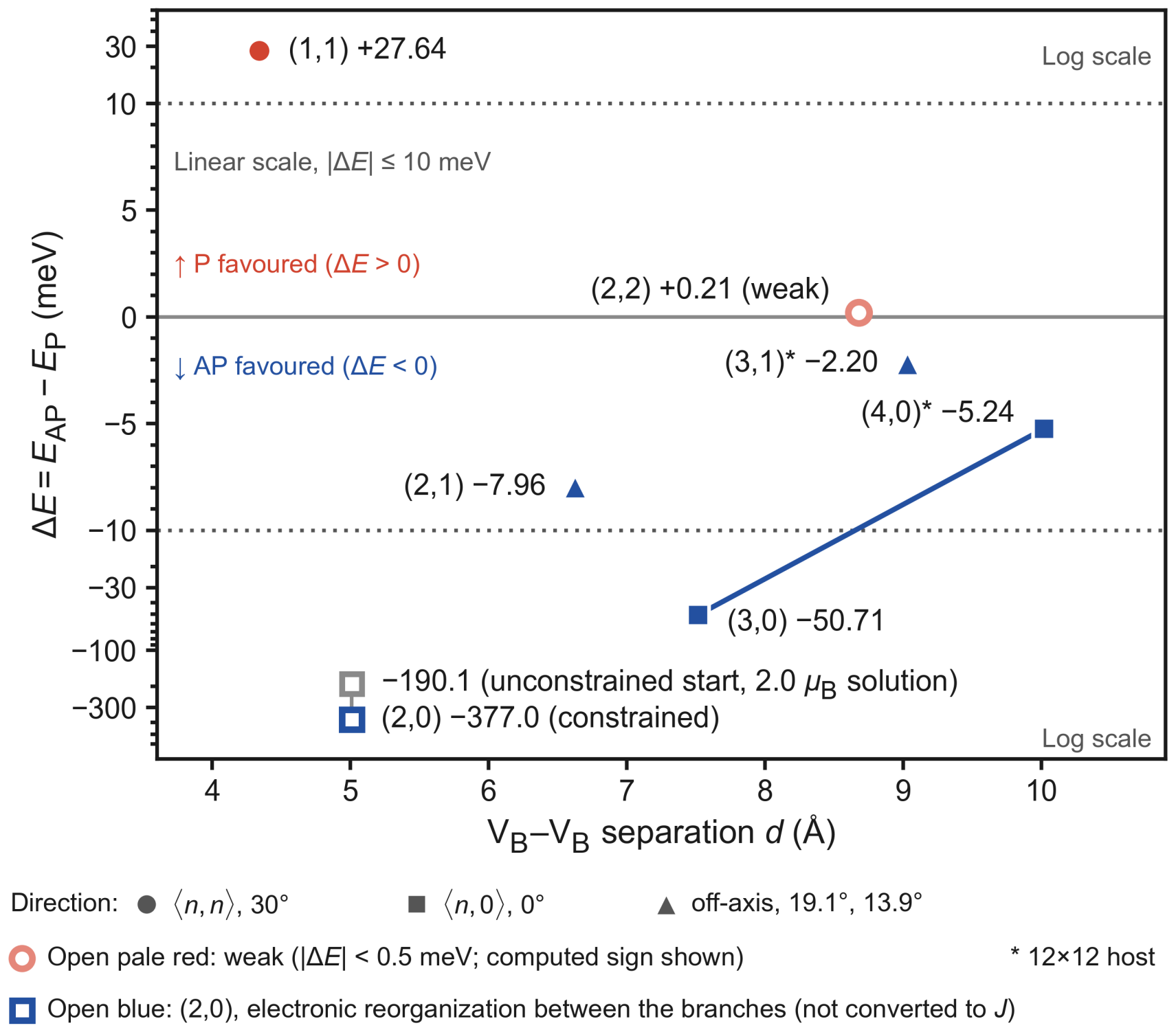

**Figure S2. Splitting versus ideal vacancy separation.** Colours identify P or AP preference and shapes the lattice-direction family; the axis is linear within ±10 meV and logarithmic outside. The weak (2,2) value is open pale red, and (2,0) is open blue because its branch difference includes electronic reorganization. The grey (2,0) marker uses the unconstrained $2\mu_B$ solution and is not mapped to an exchange parameter. The line joining the two non-weak, endpoint-consistent points (3,0) and (4,0) is a guide, not a decay fit. Source data are deposited (Section S8.3).

## S3. Relaxation and hybrid-functional checks

Neither branch-specific relaxation nor the HSE06 hybrid functional reverses the signs of the two reference pairs in the calculations performed.

### S3.1. Relaxed geometries and the four-energy comparison

P geometries are relaxed with the standard (450 eV) settings to maximum forces of 0.015–0.030 eV Å$^{-1}$. AP geometries are also relaxed for (1,1) and (3,0) in 10×10. The resulting structures remain nearly planar, with an out-of-plane range no larger than 0.007 Å. Shortest B–N bonds span 1.381–1.398 Å.

For both reference pairs we evaluate P and AP at each branch's relaxed coordinates; this separates the vertical splitting from the relaxation energy and tests whether the opposite preferences arise simply from comparing different geometries (Table S4).

**Table S4. Four-energy geometry comparison.** $R_{\mathrm{P}}$ and $R_{\mathrm{AP}}$ are the separately relaxed structures. The vertical splitting uses $R_{\mathrm{P}}$ for both branches; the relaxed splitting compares each branch at its own geometry. The AP relaxation energy is $\lambda_{\mathrm{opp}} = E_{\mathrm{AP}}(R_{\mathrm{P}}) - E_{\mathrm{AP}}(R_{\mathrm{AP}})$. Total energies are in eV and splittings in meV. All four energies use 10×10 supercells, a 520 eV cut-off and an electronic convergence criterion of $10^{-8}$ eV.

| Δ | E↑↑(R↑↑) | E↑↓(R↑↑) | E↑↓(R↑↓) | E↑↑(R↑↓) | ΔE vertical (meV) | $\lambda_{\mathrm{opp}}$ (meV) | ΔE relaxed (meV) | E↑↑(R↑↓) − E↑↑(R↑↑) (meV) | site-reversed − direct (μeV) |
|---|---|---|---|---|---|---|---|---|---|
| (1,1) | -1733.49376515 | -1733.46612900 | -1733.46765656 | -1733.49340298 | +27.64 | 1.53 | +26.11 | +0.36 | 0.000 |
| (3,0) | -1733.91466822 | -1733.96538081 | -1733.96797071 | -1733.91377818 | -50.71 | 2.59 | -53.30 | +0.89 | 0.000 |

At the P-relaxed geometry, the largest residual forces on the AP branch are 0.14–0.15 eV Å$^{-1}$ for the reference pairs. The AP relaxation energies are 1.53 meV for (1,1) and 2.59 meV for (3,0), and the relaxed-branch splittings remain +26.11 and −53.30 meV. At the AP-relaxed coordinates the vertical splittings are +25.75 and −54.19 meV; this is a different comparison, not to be interchanged with the former.

The P energies increase by 0.36 and 0.89 meV at the corresponding AP geometries. Relaxation therefore adjusts the splittings by a few meV without creating their opposite signs; this does not make geometry generally unimportant, since relative vacancy placement is itself the central structural variable.

## S3.2. Hybrid-functional results

The matched HSE06 calculations test the functional dependence at fixed geometry rather than establish an exact reference; the hybrid functional increases the vacancy projections and opens the same-spin gap at (3,0).

We use HSE06 with 25% exact exchange and screening parameter 0.2 Å$^{-1}$.[11,12] Each hybrid branch is initialized from the converged PBE orbitals at unchanged coordinates and charge and in the same supercell, with $10^{-7}$ eV electronic convergence; hybrid calculations started without PBE orbitals did not converge in the attempts made. The comparisons therefore establish the persistence of the PBE-initialized branches. The principal comparisons use 8×8 for (1,1) and (2,1), 10×10 for (3,0) and 6×6 for (2,0). In these comparisons, AP calculations with the initial vacancy moments reversed (site-reversed initializations) reproduce the original AP energies within $10^{-7}$ eV, an additional check on the branch assignment (Table S5).

**Table S5. Hybrid-functional branch energies.** The 5×5, 32% result is retained as a small-cell comparison and is not size-converged; its large site-reversed − direct difference reflects competing AP solutions. The (2,0) hybrid results are given in Section S1.5.

| Δ | super cell | exact exchange | E↑↑ (eV) | E↑↓ (eV) | ΔE (meV) | site-reversed − direct (μeV) | per-vacancy ↑↑ (μB) | per-vacancy ↑↓ (μB) |
|---|---|---|---|---|---|---|---|---|
| (1,1) | 8×8 | 25 % | -1278.705756 | -1278.653960 | +51.80 | 0.00 | +1.442/+1.483 | +1.426/-1.383 |
| (1,1) | 8×8 | 32 % | -1329.495938 | -1329.440257 | +55.68 | -0.01 | +1.466/+1.505 | +1.450/-1.407 |
| (2,1) | 8×8 | 25 % | -1278.987998 | -1278.989852 | -1.85 | 0.00 | +1.433/+1.434 | +1.433/-1.431 |
| (3,0) | 10×10 | 25 % | -2015.054370 | -2015.077733 | -23.36 | 0.02 | +1.434/+1.434 | +1.427/-1.426 |
| (1,1) | 5×5 | 32 % | -499.553030 | -499.412983 | +140.05 | -23608.56 | +1.605/+1.601 | +1.358/-1.505 |

The matched PBE-to-HSE06 changes are +24.1 meV at (1,1) and +6.0 meV at (2,1) in 8×8, and +27.3 meV at (3,0) in 10×10. At (2,0) in 6×6, the splitting of the branches continued without the spin constraint changes from −403.86 to −238.79 meV, a functional shift of 165.1 meV (165.071 meV in Table S3) that retains the AP ordering. All tested signs persist. At (1,1), the PBE state used to initialize HSE06 gives

+27.66 meV, compared with the +27.60 meV of the lowest-energy AP solution used in the supercell series.

HSE06 reverses the magnitude ranking of the two reference pairs while preserving their opposite preferences. Increasing the exact-exchange fraction at (1,1) from 25% to 32% raises the splitting further, from +51.80 to +55.68 meV. These comparisons quantify functional sensitivity; they do not provide error bars on the PBE or hybrid values.

Because the functional shifts are large compared with the sub-meV numerical controls, sign persistence is the principal cross-functional result. No HSE06 values are available for (2,2), (3,1) and (4,0); in particular, the two small AP splittings in 12×12 should not inherit the robustness established for the reference pairs.

## S4. Electronic structure at fixed nuclei

We compare P and AP at identical nuclei to locate where charge, spin magnitude and orbital levels differ. These quantities are descriptors: each can show a reproducible electronic feature without defining its contribution to the self-consistent total energy.

We distinguish PAW-projected moments, which are obtained from the lm-resolved PAW projector functions rather than by integrating the spin density within chosen atomic spheres; real-space integrals of the spin density over spatial partitions; and the constrained sphere moments of Section S5. The three are not interchangeable.

### S4.1. Vacancy-projected frontier levels

The isolated charged vacancy has two spin-active in-plane dangling-bond states, consistent with the $e'$ assignment of Ivády and co-workers.[13] The pair has four frontier σ levels per spin channel. In P, their occupations are four majority and zero minority electrons; in AP, each spin channel contains two occupied and two empty frontier levels (Figure S3). This provides the occupation-based motivation for an AP-favouring occupied–empty mixing contribution.

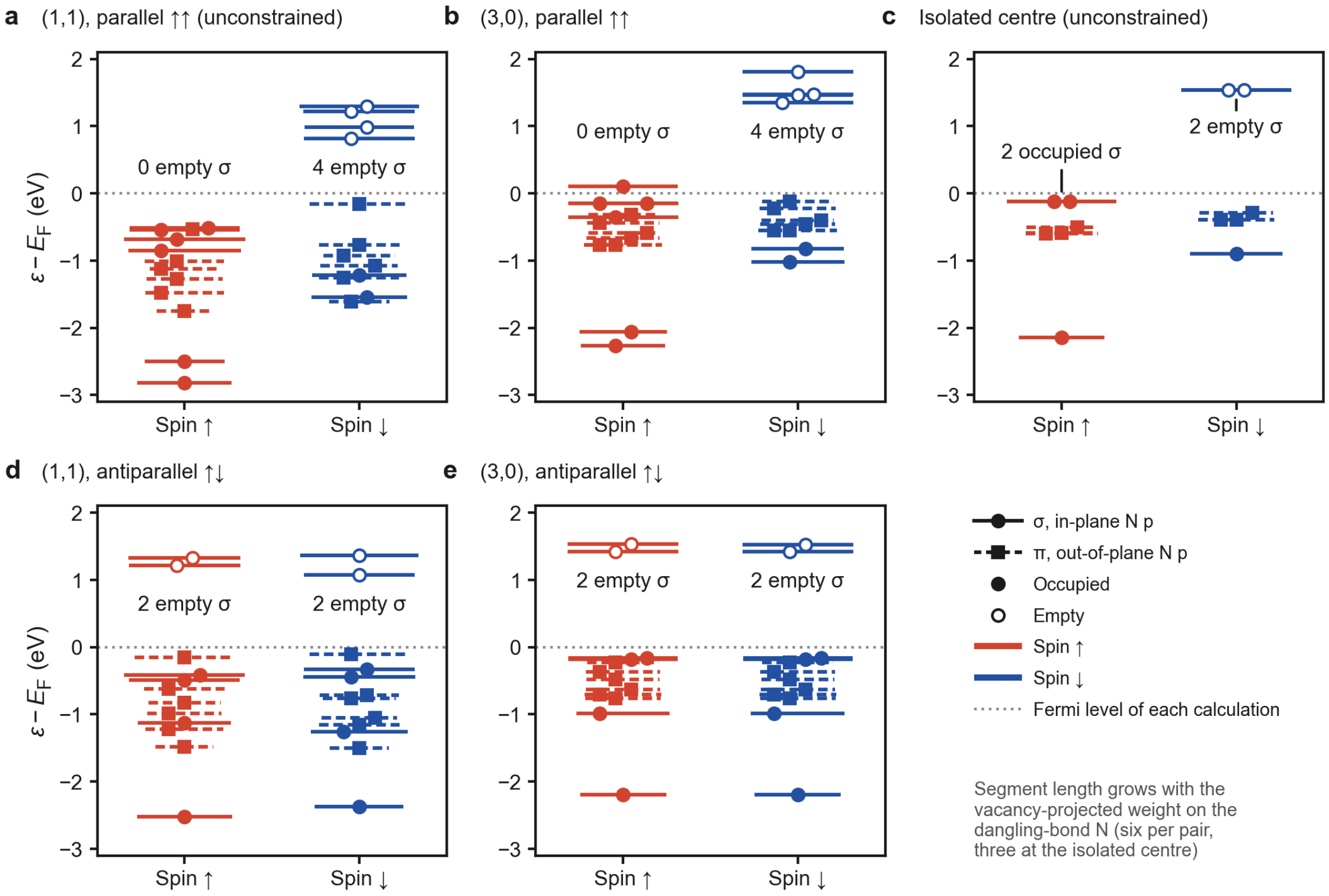

**Figure S3. Vacancy-projected frontier levels.** Panels compare (1,1), (3,0) and an isolated vacancy in 10×10. Red and blue denote spin up and down, circles and squares in-plane σ and out-of-plane π character, and filled and open symbols occupied and empty levels; segment length reflects vacancy-projected weight. Each panel uses its own Fermi-energy reference, so level positions are not aligned across panels; the occupation counts, not level offsets, establish the four/zero pattern in P and the two/two pattern in AP. Source data are deposited (Section S8.3).

## S4.2. Definitions and numerical floors

**Local moments and densities.** PAW-projected moments and orbital weights are obtained by projecting the Kohn–Sham orbitals onto the atom-centred, lm-resolved PAW projector functions. Vacancy moments sum the projections on their three neighbouring nitrogen atoms; they are approximate local descriptors, not sphere-integrated moments, atomic charges or quantum spin assignments. Real-space analyses use the densities on the FFT grid of the plane-wave calculation, comprising the pseudo-valence density and the PAW compensation charge rather than the full all-electron density.

**Regions.** Each vacancy's three dangling-bond nitrogens form its triad. Partition P1, the primary partition, uses nearest-atom Voronoi cells within $|z - z_0| \leq 2.5$ Å, and partition P2, used as a sensitivity test, radical Voronoi cells with atomic radii; P1 and P2 label spatial partitions, not the P branch. At (1,1) we distinguish contact N (N99, N100, N109), bridging B (B1, B10), non-facing N (N108, N189, N110) and the contact hollow, a radius-0.70 Å cylinder through the intervening ring; at (3,0), the chain ends (N99, N119) and the interior (B10, N109, B20).

**Density contrasts.** For region $\Omega$, $Q_\Omega = 1/2 \int_\Omega |n_{\mathrm{AP}} - n_{\mathrm{P}}|\, dV$ measures charge-density redistribution, not net transferred charge. The spin-magnitude contrast is $C_A = \int_\Omega |m_{\mathrm{AP}}|\, dV - \int_\Omega |m_{\mathrm{P}}|\, dV$, and the squared-density contrast is $C_S = \int_\Omega \left(m_{\mathrm{AP}}^2 - m_{\mathrm{P}}^2\right)\, dV$. Here $m = n_\uparrow - n_\downarrow$.

**Rendering of Figure 1c,d.** Figure 1c,d of the main text uses the unsmoothed spin density on this grid, $m = n_{\uparrow} - n_{\downarrow}$. Views are orthographic, 52° from the sheet normal, with identical scale, isovalue and lighting; scale bars apply horizontally.

**Facing atoms.** The facing set comprises triad nitrogens in inter-triad N–N pairs within 0.10 Å of the shortest such distance; its summed $C_A$ is $dA_F$. It is the three-nitrogen contact at (1,1) and the two chain ends at (3,0).

**Spectral reference and classification.** Within each placement, Γ-point levels are aligned to the mean core potential of remote host atoms, those at least six bonds from a triad nitrogen and more than two bonds from a displaced atom. PAW weights are magnitudes, not phases. A state is π when its out-of-plane fraction is at least 0.9 and σ when it is at most 0.1.

**Contact-localized $\pi$ state.** At (1,1), the selected occupied π band has the largest contact-N $p_z$ weight, at least 0.05 on each contact nitrogen and at least 0.30 eV separation from the next relevant π level. Where no state qualifies, the highest $\pi$ defect level and the $\pi$ state of largest facing weight are reported separately, not as the contact state.

**Frontier subspaces.** The four highest σ bands with triad $(s + p_x + p_y)$ share of at least 0.30 within 2.5 eV of the highest occupied state define the $e'$ manifold; the next two eligible levels above −10 eV define the $a_1{}'$ manifold. Both are operational assignments, not exact symmetry labels.

**Occupied–empty descriptors.** In AP, $p_{\mathrm{oe}}$ is the partner-triad share of the occupied same-spin $e'$ weight, evaluated in each spin channel (the single values quoted in Section S4.4 are means over the two spin channels), and $q_{\mathrm{oe}}$ the empty-subspace share on the opposite triad, with centroid gap $G_{\mathrm{oe}}$. The proxy $p \times G = \sum_s p_{\mathrm{oe},s}\, G_{\mathrm{oe},s}$ combines this admixture with the occupied–empty centroid gap in the sum $\sum_s p_{\mathrm{oe},s}\, G_{\mathrm{oe},s}$ over the two spin channels $s$. It follows the scale of second-order mixing in a fixed two-level model; it is not a DFT energy and is zero in P by the occupation convention of the selected manifold. Its restriction to the mirror-odd component at (1,1) is exploratory.

**Effective two-level coupling.** For a mixed same-spin pair in P with splitting $\Delta\varepsilon$, $t_{\mathrm{2L}} = \Delta\varepsilon\sqrt{w(1-w)}$, where $w$ is the normalized partner-vacancy weight; the P-branch values in the two spin channels are denoted $t_{\uparrow}$ and $t_{\downarrow}$. Equal onsite energies give $w = 1/2$ and $t_{\mathrm{2L}} = \Delta\varepsilon/2$. It reproduces the selected two-level splitting and mixing but is not a many-body hopping integral. These two-level and occupied–empty quantities describe the Kohn–Sham states and are not separately identified contributions to $\Delta E$.

**Responses and thresholds.** The full-span response is $D_{\pm}X = X(+\delta) - X(-\delta)$; a half-span percentage divides by twice the undisplaced magnitude, a full-span percentage does not. A descriptor "tracks" the energy when it keeps the specified sign relation to it across the tested modes; agreement under two modes does not establish a general functional relation. Thresholds combine rounding and repeat-calculation noise. Correlation under a displacement is not an energy decomposition or proof of causation.

**Electronic acceptance.** Accepted P/AP pairs keep the expected total moment, frontier pattern and numbers of occupied spin-up/spin-down states: 400/396 for P and 398/398 for AP in 10×10, and 576/572 and 574/574 in 12×12.

**Density checks.** The 22 analysed densities integrate to 796 electrons within $2.6 \times 10^{-7}$ electron; total moments reproduce P at 4 μB and AP within $5 \times 10^{-9} \mu_{\mathrm{B}}$. Partitions close on the full cell within $5 \times 10^{-12}$, and identical inputs give zero contrast.

**Repeat floors.** Same-branch repeats give whole-cell $Q$ no larger than $1.8 \times 10^{-4}$ electron and regional $|C_A|$ floors of approximately $10^{-7}$–$10^{-5} \mu_{\mathrm{B}}$. Repeated eigenvalues agree within $7 \times 10^{-5}$ eV, or $7.7 \times$

$10^{-4}$ eV for (1,1) AP at its relaxed geometry. States are tracked across distortions through projected magnitudes (weight-vector similarity at least 0.9997), because overlaps between displaced and undisplaced wavefunctions were not computed. These are repeat sensitivities and numerical checks, not measures of physical accuracy.

## S4.3. Charge and spin redistribution between the alignments

At fixed P-relaxed nuclei (Figure S4), the charge redistribution is small compared with the total valence density: $Q = 0.021$ electron for (1,1) and 0.032 for (3,0), of which approximately 58–64% lies outside the six dangling-bond nitrogens in P1 and 62–66% in P2. The opposite alignment preferences are therefore not accompanied by a large, spatially isolated charge transfer.

The AP branch has less total spin magnitude by 0.227 μB at (1,1) and 0.127 μB at (3,0). For (1,1), much of the change occurs on vacancy B's two contact nitrogens. For (3,0), the chain-centre N109 carries a PAW-projected moment of 0.058 μB in P and is unpolarized in AP.

The P-branch two-level couplings are 0.229/0.231 eV for (3,0) and 0.168/0.146 eV for (1,1), giving a squared ratio of 0.40–0.54. They are distinct from the Wannier-derived hopping blocks of Section S8.1, although both rank the two configurations in the same order. Only (1,1) has the selected occupied $\pi$ state localized at the compact contact; its splitting falls from 0.379 to 0.045 eV in AP.

A smaller local spin magnitude need not imply a different local quantum spin. In an orbital-tail example, the N99-centred $e'$ state reaches onto the partner triad with similar weight in P and AP; its spin contribution adds to the partner's in P and opposes it in AP, so the projected contact spin can decrease through overlap cancellation. This example motivates, but does not uniquely decompose, the self-consistent density contrast.

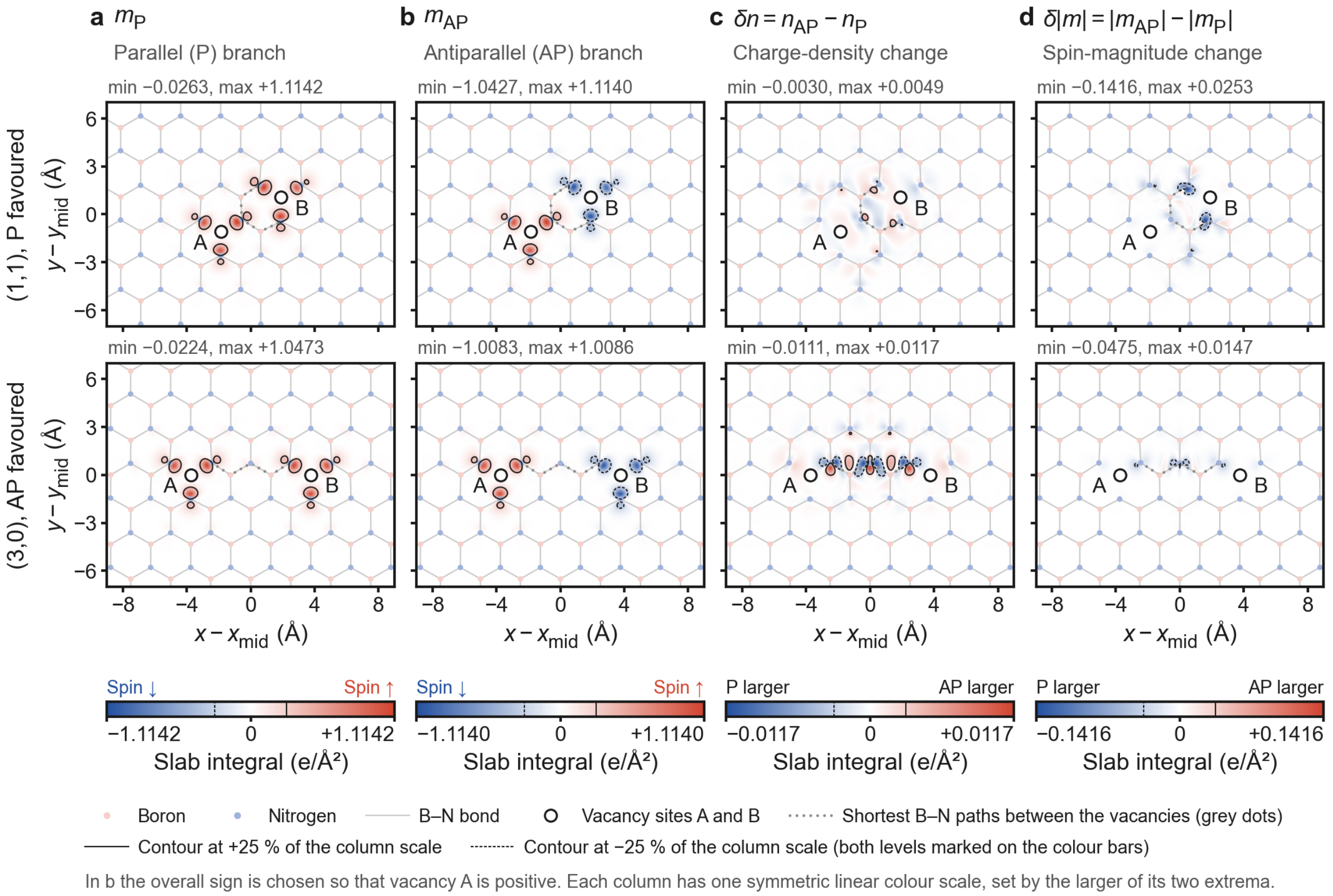

**Figure S4. Charge and spin redistribution at fixed nuclei.** Columns show P spin density, AP spin density, charge-density change and spin-magnitude change for (1,1) and (3,0). Maps are unsmoothed slab integrals over $|z - z_0| \leq 2.5$ Å. Each column uses one symmetric colour scale with ±25% contours and printed extrema. Vacancies A and B are open rings, B–N bonds grey and shortest inter-triad paths dotted; the AP sign makes vacancy A positive. Source data are deposited (Section S8.3).

## S4.4. The (1,1) contact

The compact contact accounts for most of the (1,1) spin-magnitude difference; Table S6 lists real-space integrals, PAW-projected moments and spectral quantities separately because their meanings differ.

**Table S6. Electronic contrast at the (1,1) contact.** Values compare P and AP at identical P-relaxed coordinates in 10×10 PBE. Local-majority-signed projections are positive along the majority direction of the corresponding vacancy, allowing moment magnitudes to be compared after reversal. Dashes identify entries not tabulated, not zeros.

| quantity | P | AP | ΔX = AP − P |
|---|---|---|---|
| $C_A$, contact-N atom cells | — | — | −0.140 μB (62 % of the whole-cell −0.227 μB) |
| $C_A$, intermediate host shell / non-facing N / contact hollow | — | — | −0.066 (29 %) / −0.005 / +0.0026 μB |
| PAW-projected contact-N moment (local-majority-signed) | 1.529 μB | 1.378 μB | −0.151 μB, of which −0.122 μB in-plane (px + py) |
| PAW-projected non-facing-N moment | 1.227 μB | 1.225 μB | ≈ 0 |
| contact π state: band (↑/↓), occupation | 399/396, filled | 398/398, filled | filled in all four channels |
| contact π state: spin splitting | 0.3787 eV | 0.0453 eV | −0.333 eV |
| contact π state: midpoint | — | — | +17.5 meV (P lower; changes by < 0.03 meV with the reference) |
| filled π pair's own contact polarization | 0.0056 μB | 0.0022 μB | — |
| total contact-N pz moment (all occupied states) | 0.074 μB | 0.067 μB | — |

The three contact-nitrogen cells contribute −0.140 μB, 62% of the whole-cell contrast. The contact hollow instead gives +0.0026 μB. About 77% of the contact $\pi$ state's captured projector weight is on the three contact $p_z$ orbitals, corresponding to approximately 46% of the full band norm. Its 17.5 meV AP-minus-P midpoint shift reports a level response, not the energy of a contact interaction.

A positive AP-minus-P $\pi$ midpoint also occurs at (3,0), where AP is energetically lower. The midpoint alone therefore cannot determine the preferred alignment. Likewise, overlapping same-sign profiles satisfy $|m_A - m_B| \leq |m_A + m_B|$, so reversing one contribution can reduce local spin magnitude without establishing a separate polarization-energy gain.

At (3,0), the facing contrast is −0.052 μB, 2.7 times smaller than at (1,1), even though its PBE energy splitting is larger, a magnitude ranking that HSE06 reverses (Section S3.2). The spin-averaged AP occupied–empty admixture (Section S4.2) is larger at (1,1) than at (3,0), 0.047 versus 0.023 in 10×10 PBE at P-relaxed coordinates (Figure S5c shows the two spin channels separately), whereas the squared P-branch two-level coupling at (1,1) is only 0.40–0.54 of that at (3,0) (Section S4.3). Different descriptors therefore do not provide a unique ranking of energetic contributions. This disagreement between descriptors is retained as a limitation of a single-channel interpretation.

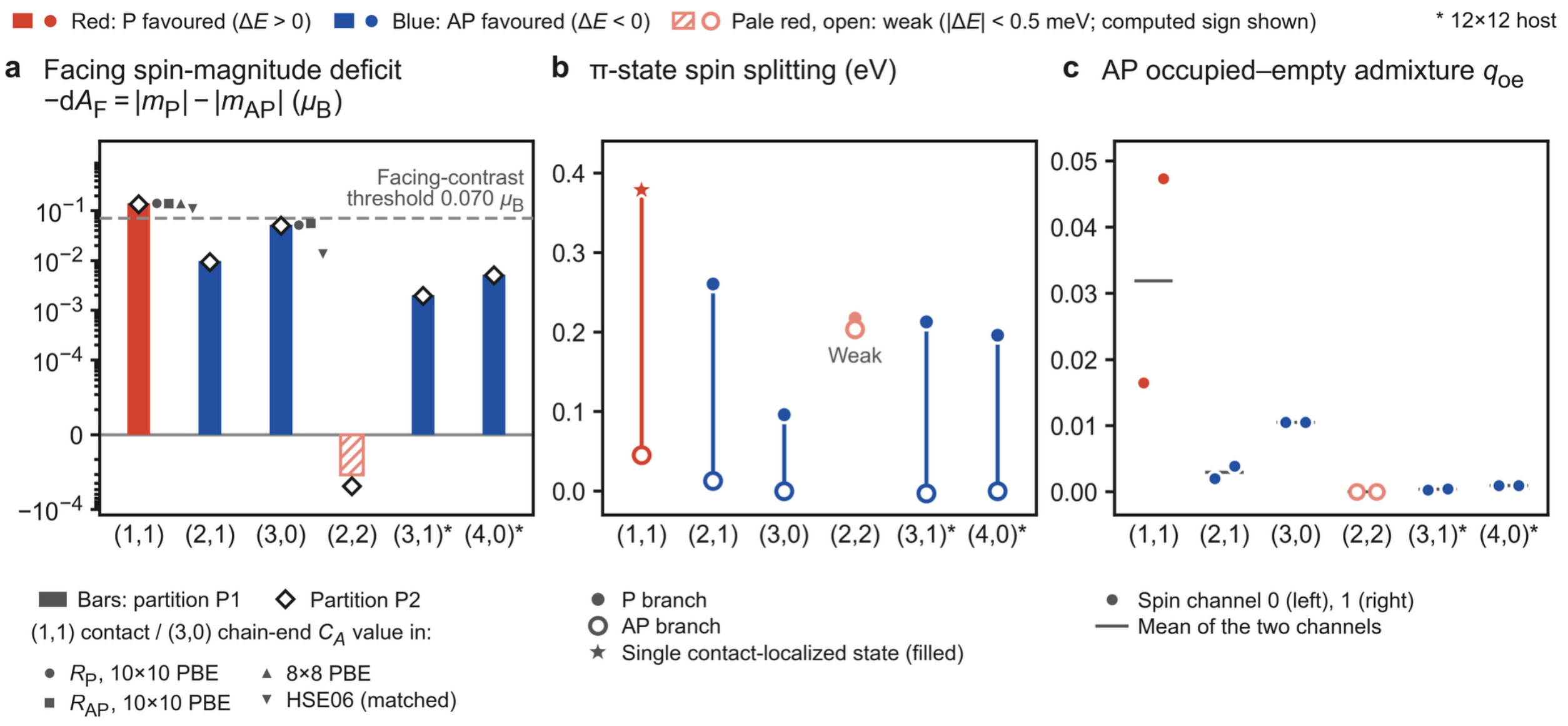

**Figure S5. Baseline electronic signatures across placements. (a)** Facing-atom spin-magnitude deficit with both partitions and the 0.070 μB threshold of the transfer test (Section S6.6); smaller symbols are reference-geometry and functional comparisons. **(b)** Top-$\pi$ splittings in P and AP, with a star identifying the single contact-localized state at (1,1). **(c)** AP occupied–empty admixture by spin channel; channels 0 and 1 denote spin up and spin down, not spin-0 and spin-1 states. Colours denote the sign of the computed branch difference, with weak and larger-cell cases identified. These are electronic descriptors rather than energy terms. Source data are deposited (Section S8.3).

# S5. Constrained noncollinear calculations

A rigid, bilinearly coupled pair predicts $E(\theta) = c_0 + c_1 \cos\theta$ and fixed local moments; we test this at interior angles. The atom-wise-constrained PBE energies of (1,1) depart reproducibly from the endpoint interpolation, whereas one interior angle of (3,0) is compatible with it. The tests concern constrained Kohn–Sham states, not spin-pure multiplets; uncompleted checks and retrospective amendments are reported with each test.

## S5.1. Protocol and endpoint agreement

**Numerical settings.** We use noncollinear PBE without spin–orbit coupling, a 520 eV plane-wave cut-off, Γ sampling, the blocked Davidson algorithm, real-space projector evaluation and electronic convergence to $10^{-7}$ eV. The 8×8 and 10×10 cells contain 508 and 796 electrons in 560 and 848 bands, twice the number of bands used in the corresponding collinear calculations, at fixed P-relaxed collinear coordinates. Further noncollinear settings are given in the reproducibility record.

**Supercells.** The completed sets use the P-relaxed (1,1) structure in 8×8 and the P-relaxed (3,0) structure in 10×10, so they do not isolate a placement-only difference. The larger (1,1) cell lacks a converged 180° endpoint, so its completed 0° and near-90° calculations enter only the descriptive comparison of Section S5.4.

**Constraint.** Following Ma and Dudarev,[14] a penalty $E_p = \lambda \sum_i (|\mathbf{M}_i| - \hat{\mathbf{e}}_i \cdot \mathbf{M}_i)$ acts on the six dangling-bond nitrogen moments, obtained as weighted integrals within atomic spheres (radii 0.741 Å for N and 0.600 Å for B) rather than PAW-projected; the nitrogen spheres capture approximately 0.39 of the PAW moment. Target directions are along $z$ for vacancy A and at angle $\theta$ to $z$ for vacancy B: A targets $+z$, and B targets $(\sin\theta, 0, \cos\theta)$, at sequential strengths of 5, 10 and 20 eV $\mu_B^{-1}$. Each atomic direction is

constrained separately, which restricts internal relaxation at a fixed collective vacancy direction. Each strength stage is initialized from the converged state of the previous one; we call such a sequence a path.

**Energy and angle.** We subtract the penalty once from the zero-smearing energy: $E_{\mathrm{phys}} = E(\sigma \to 0) - E_p$. The reported angle is that between the summed sphere moments of the two vacancies; the finite precision of the recorded moments gives it a rounding interval of approximately ±0.33°. Angles refined from the constraint fields and from the PAW projections are reported separately.

**Two residuals.** The bilinear residual subtracts the straight line in $\cos\theta$ joining the endpoints computed in the same supercell; the three-term residual subtracts $c_0 + c_1\cos\theta + c_2\cos^2\theta$, with coefficients fixed by the endpoints and near-90° state. A failed three-term prediction is a different test from a nonzero bilinear residual.

**Chronology and checks.** Angular rules were fixed before the outputs they evaluated, except that the near-112.5° point was added after the near-135° test failed, the global-rotation criterion at 90° followed inspection of the 0° result, and an auxiliary acceptance criterion for the larger cell was loosened after inspection.

The noncollinear endpoints reproduce the collinear splittings computed in the same supercell within 0.127 meV at (1,1) and 0.074 meV at (3,0). The penalty is effectively inactive there: for (1,1), constrained and penalty-free energies differ by less than 0.0004 meV, and the 0° energy is unchanged within $3 \times 10^{-9}$ meV between strengths 5 and 10. The accepted endpoint gap is 27.7253 meV; penalty-free calculations may retain slight canting, so their angles are reported rather than assumed collinear. The endpoint agreement compares collinear and noncollinear calculations with different numerical settings; it is not a convergence series in all parameters, and the planned strength-20 endpoint check was not completed.

## S5.2. Angular energies of (1,1) and the held-out tests

Table S7 lists the constrained (1,1) states.

**Table S7. Constrained angular states of (1,1) in 8×8.** Energies are relative to $E_{\mathrm{phys}}(0°)$ and residuals refer to the bilinear endpoint interpolation. Refined values are in parentheses; dashes mark unreported or inapplicable entries. Zero-angle results at strengths 5 and 10 coincide to the precision stated in Section S5.1.

| θ requested | λ (eV/μB) | θ achieved: sphere-moment (refined) | $E_{\mathrm{phys}} - E_{\mathrm{phys}}(0°)$ (meV) | $E_p$ (meV) | r: sphere-moment (refined) (meV) | PAW-projected moment A/B (μB) |
|---|---|---|---|---|---|---|
| 0° | 5 and 10 | 0.000° | 0 | $< 10^{-8}$ | — | 1.355/1.395 |
| 45° | 10 | 44.556° | +6.0375 | 0.161 | +2.053 | — |
| 45° | 20 | 44.891° (44.854°) | +6.1975 | 0.089 | +2.156 (+2.162) | 1.350/1.378 |
| 90° | 5 | 89.339° (89.3515°) | +17.5733 | 0.427 | +3.871 (+3.8675) | — |
| 90° | 10 | 89.669° (89.678°) | +17.9906 | 0.248 | +4.208 (+4.206) | — |
| 90° | 20 | 89.890° (89.839°) | +18.2382 | 0.135 | +4.402 (+4.414) | 1.339/1.338 |
| 112.5° | 10 | 112.405° (112.255°) | +23.2892 | 0.177 | +4.143 (+4.176) | — |
| 112.5° | 20 | 112.294° (112.379°) | +23.4664 | 0.096 | +4.345 (+4.326) | 1.334/1.317 |
| 135° | 10 | 134.889° | +26.8307 | 0.079 | +3.185 | — |
| 135° | 20 | 134.889° (134.937°) | +26.9091 | 0.043 | +3.263 (+3.255) | 1.328/1.295 |
| 180° | 5 | 180.000° | +27.7253 | $< 10^{-8}$ | — | 1.325/1.276 |

The near-90° path is initialized from the P state, and each other interior angle is computed along a single path initialized from the strength-10 near-90° state; three independently initialized near-90° calculations at strength 10 agree within 0.0008 meV, which shows reproducibility but not uniqueness.

**Finite-strength sensitivity.** The near-90° energy rises by 0.42 meV from strength 5 to 10 and 0.25 meV from 10 to 20, partly because the achieved angle moves; the last step meets the 0.5 meV criterion but is not an asymptotic bound. Model extrapolations to infinite constraint strength give near-90° energies of 18.45–18.60 meV and residuals of 4.58–4.75 meV. The strength-20 penalty, 0.135 meV, exceeds the 0.1 meV target, and the planned second path at strength 20 was not completed.

**Rotation and continuity.** A global rotation changes the energy by approximately −0.00044 meV at 0° and −0.00052 meV near 90°, as expected without spin–orbit coupling. Occupations stay integer with no observed frontier crossing, the gap rises from 0.988 to 1.204 eV, and triad populations remain within 0.002 electron.

**Local moments.** Vacancy B's PAW-projected moment decreases from 1.395 to 1.276 μB along the path, an 8.9% range relative to its mean, and fails the predefined 5% rigidity test; A varies by 2.2%. Sphere moments give the same result, with a 9.4% variation.

**Internal constraint cost.** At interior angles, 79–93% of the penalty maintains alignment within individual nitrogen triads, so the physical energy may include a cost of suppressed intra-vacancy relaxation; coupling across the contact may also contribute, and we do not separate the two.

**Higher-order terms.** The three-term coefficient ratio $|c_2/c_1|$ is approximately 0.32, above the predefined 0.10 threshold for a small higher-order term. Higher-degree fits are not independent tests, and the residuals form a broad positive excess between 90° and 112.5° rather than identifying a unique microscopic interaction.

**Held-out three-term test.** Before any 45° or 135° calculation, we fixed the coefficients at $(18.264845, -13.862650, -4.402195)$ meV from the endpoint and near-90° strength-20 energies and required absolute errors no larger than 0.5 meV at both achieved angles. The errors are −0.037 meV near 45° and +1.053 meV near 135°, so the three-term form fails its joint criterion, also over the angle-rounding interval. The independent evidence against the bilinear interpolation is the positive 2.2–4.4 meV interior residual, not this failure alone.

**Additional angle.** A near-112.5° point, added after this failure, was compared with three continuations of the angular curve beyond 90°, each specified before the point was calculated: an excess localized near 135° (the localized form), a smooth higher-harmonic continuation evaluated with a quartic form (the smooth form) and a change of slope at 90° (the kink-like form). Its deviations from them are +0.576, −0.176 and +0.052 meV. The predefined rule therefore retains smooth or kink-like behaviour, but the localized alternative exceeds its threshold by only 0.076 meV, less than the rule's 0.25 meV tolerance, so this point does not discriminate decisively between the shapes.

**Constraint-radius test.** An alternative sphere definition, N 0.900 Å and B 0.450 Å, captures approximately 0.54 rather than 0.39 of the PAW moment and changes endpoint energies negligibly; being atom-wise, it tests radius sensitivity, not a collective-vacancy constraint. The radius change alters the bilinear residual by at most 0.16 meV and the three-term error by at most 0.03 meV, within the ±0.3 meV criterion; this shows stability under the tested change, not radius independence. The unmet or uncompleted targets are the B-moment rigidity test, the preferred near-90° penalty, and the second path and endpoint check at strength 20.

## S5.3. One interior angle for (3,0)

For (3,0), a single near-90° angle was calculated, at three strengths and with two paths at strength 5. At the refined angle, the residuals are −0.571 and −0.568 meV at strength 5 (two paths), −0.457 meV at 10 and −0.385 meV at 20 (−0.408 and −0.409 meV at 10 and 20 for the sphere-moment angle). Every reported residual lies within 0.6 meV of the bilinear interpolation, and vacancy projections vary by approximately 2%. With no held-out angle or prospective shape rule, one point cannot validate the curve or establish a singlet.

The last strength increment changes the energy by 0.195 meV; the strength-5 angle misses the preferred 1° target, and the strength-20 penalty is 0.1018 meV. Two paths at strength 5 agree within 0.0025 meV, and higher strengths have one path. A penalty-free AP endpoint agrees closely with its constrained counterpart; no penalty-free P endpoint was calculated.

### S5.4. An incomplete (1,1) control in the 10×10 supercell

This set was designed to test whether the near-90° bilinear residual agrees with the 8×8 value within 0.5 meV. Three attempts to converge the AP endpoint failed for reasons we have not identified, so no unconverged energy is used and no conclusion on the supercell dependence of the residual is drawn; the collinear 8×8-to-10×10 comparison is unaffected. With matched settings at strength 20, the near-90° energy relative to 0°, $\Delta E_{90}$, is 18.238 meV in 8×8 and 18.251 meV in 10×10; at strength 10, it is 17.991 and 18.002 meV. Compensating changes in the endpoint gap and nonlinear contribution are possible, so no larger-cell $r(90°)$ or higher-order coefficient is inferred.

The 10×10 near-90° state has one path. An auxiliary acceptance criterion on the zero-angle constraint field was loosened after the energies had been inspected; this amendment changes no reported energy.

### S5.5. Scope and energy scales

The scope is set by PBE, Γ sampling, no spin–orbit coupling, finite strengths, atom-wise constraints and limited angles. Attributing the angular contrast to placement requires a matched-cell comparison; collective-vacancy constraints and noncollinear hybrid calculations would extend the test. Local quantum spins, multiplet energies and a unique mechanism are not established, and the accepted collinear alignment map is not overturned.

Within the conditional isotropic two-spin-1 model, $J = \Delta E/2$ and degeneracy-weighted Boltzmann factors determine thermal populations. At 300 K, the PBE (1,1) populations are approximately 80.2% quintet, 16.5% triplet and 3.2% singlet. For PBE (3,0), they are 41.9% singlet, 47.1% triplet and 11.0% quintet; using its HSE06 endpoint parameter gives 23.8%, 45.5% and 30.7%, respectively. Rounded percentages may not sum exactly to 100%. These values illustrate the energy scale under the stated model, especially at (1,1), where the angular results do not validate the bilinear assumption. The populations assume the isotropic bilinear model, fixed exchange and thermal equilibrium, and do not include optical pumping, spin–phonon relaxation or structural reorganization.

The isolated-centre zero-field splitting, approximately 3.5 GHz or 14 μeV, and point-dipole prefactor at 4.34 Å, approximately 2.6 μeV, are much smaller than the assigned electronic branch preferences. The conversion 1 meV ≈ 242 GHz is dimensional only and is not a predicted resonance. Pair-specific anisotropy, hyperfine interactions and optical selection rules were not calculated.

## S6. Deformation tests

Baseline (fixed-nuclei) and response (displacement) analyses are separate questions: a descriptor that tracks a response need not explain the baseline, and an electronic signature is not an isolated energy contribution. We examine seven modes: the (3,0) chain with off-path and chain-end controls, and the (1,1) bridge, off-path and two contact-N modes. "Partly supported" means agreement with some discriminating observations while others limit the scope, and "candidate" means plausible but not decisively tested; neither label implies a calculated microscopic energy term.

## S6.1. Connectivity, mode definitions and acceptance

Beyond the existing P/AP states, we use twelve single-point calculations for the original displaced geometries, four for the off-path control and twelve later ones for the two contact-N modes and the chain-end control, each specified before its own results.

All energy contrasts use $\Delta E = E_{\mathrm{AP}} - E_{\mathrm{P}}$, and electronic contrasts are AP minus P. We weight controlled energy responses most, then transfer to other placements, then bond and subspace descriptors, an order fixed after the contact-$\pi$ descriptor had been inspected. Internal checks reuse the same calculations and are not independent validation.

At AP-relaxed coordinates the reference splittings are +25.75 meV for (1,1) and −54.19 meV for (3,0), with the same signs as at P-relaxed coordinates; the four-energy cycles close within $4 \times 10^{-8}$ eV.

The compact (1,1) connection consists of N99–B1–N100 and N99–B10–N109 bridges. The unique shortest (3,0) chain is N99–B10–N109–B20–N119. Their shortest inter-triad N–N distances are 2.457 and 4.889 Å, respectively (Figure S6a,b).

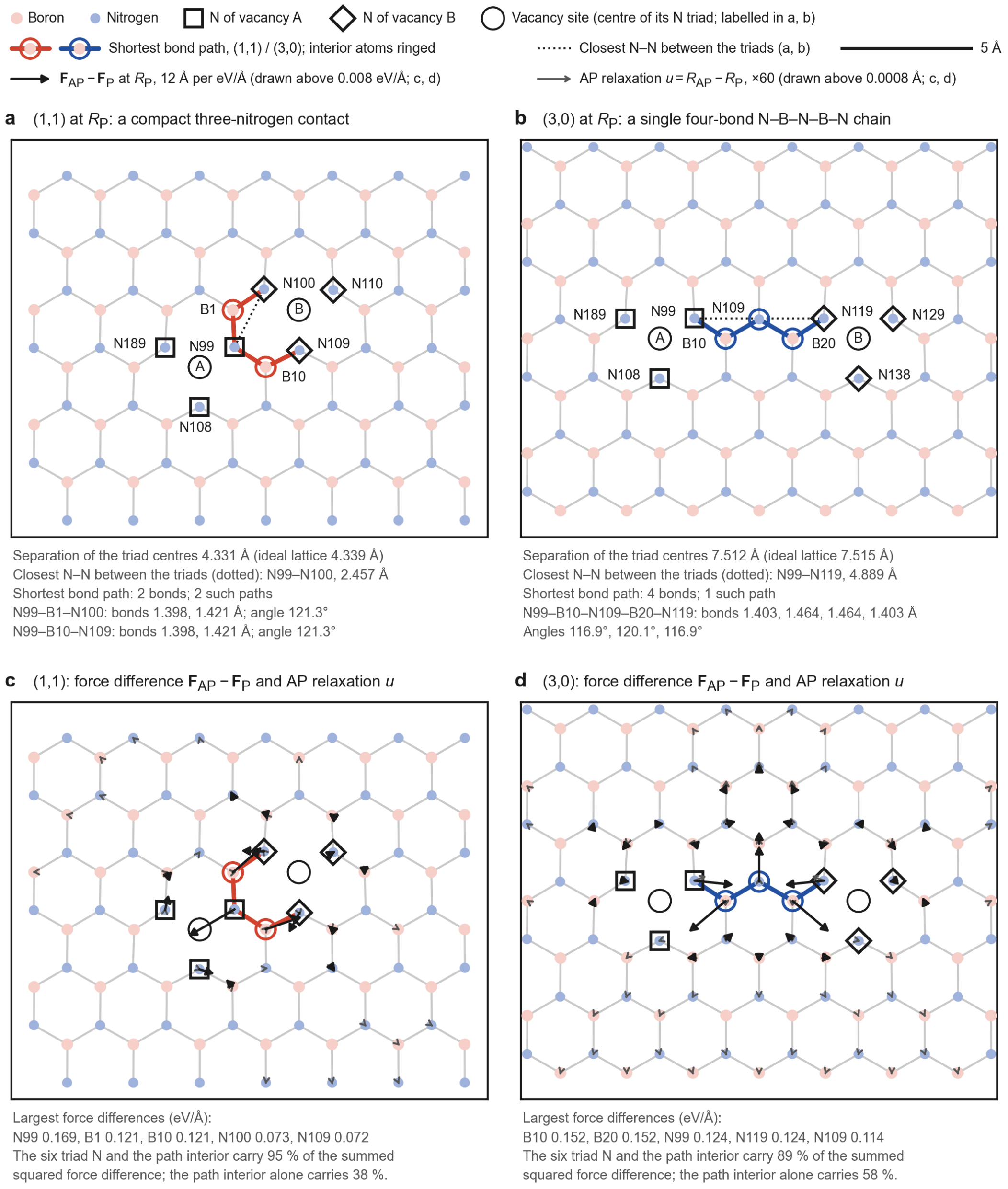

**Figure S6. Connectivity and force sensitivity of the reference pairs. (a,b)** Structures of (1,1) and (3,0), with vacancy-A nitrogens marked by squares and vacancy-B nitrogens by diamonds. Coloured lines identify shortest B–N paths and dotted lines the nearest inter-triad N–N distances. **(c,d)** P/AP force differences at P-relaxed nuclei and the AP relaxation displacements. The force difference equals $-\partial \Delta E / \partial \mathbf{R}_i$ for these states with negligible electronic entropy. Percentages beneath the panels refer to the named atom groups. Source data include the coordinates and forces.

Coordinates follow $\mathbf{R}_i(\pm\delta) = \mathbf{R}_i(0) \pm \delta \mathbf{u}_i$, with in-plane directions normalized so that the largest atomic displacement is $\delta = 0.02$ Å; unselected atoms and lattice vectors remain fixed. Table S8 defines the three original modes; the chain-end control and the contact-N modes are defined in Sections S6.4 and S6.5.

**Table S8. Original displacement modes.** Atom indices refer to each placement's own 10×10 structure. Positive displacement shortens the specified bonds but also changes other bonds and angles; all three modes are composite perturbations, not changes of a single hopping parameter.

| mode | atoms moved | direction | on an inter-vacancy path? |
|---|---|---|---|
| (1,1) bridge | the two bridge borons B1 and B10 | along the bisector of each boron's two bonds to the contact nitrogens (N99, N100 and N99, N109) | yes: the interiors of N99–B1–N100 and N99–B10–N109 |
| (3,0) chain | the chain interior B10, N109 and B20 | flattens the in-plane zig-zag (maximum atomic displacement δ); the four chain bonds shorten by different amounts and other bonds and angles change, so the mode is a composite deformation, not the variation of a single hopping integral | yes: the interior of N99–B10–N109–B20–N119 |
| (1,1) control | B11 and B20 | the bridge distortion translated rigidly by the pair vector, next to the far nitrogen N110 of vacancy B | no |

Acceptance of the displaced states requires convergence, integer occupations, a gap above 0.5 eV, the expected total and local-moment signs, triad-moment changes within 0.05 μB, retained frontier occupations and unchanged FFT grids; all twelve original displaced states pass without a branch switch. The displaced states of the later modes likewise meet their criteria and retain their branch character (Sections S6.4 and S6.5).

The finite-difference slope is $D_{\mathrm{FD}} = [\Delta E(+\delta) - \Delta E(-\delta)]/(2\delta)$; its first-order force reference is $D_{\mathrm{force}} = -\sum_i \left(\mathbf{F}_{\mathrm{AP},i} - \mathbf{F}_{\mathrm{P},i}\right) \cdot \mathbf{u}_i$. The force slopes were known when the modes were chosen, so their agreement is a consistency check and the prospective content lies in the descriptor responses. The half-span energy response is $\eta = [|\Delta E(+\delta)| - |\Delta E(-\delta)|]/[2|\Delta E(0)|]$. Percentages use this half-span normalization unless explicitly labelled otherwise. The finite-difference slopes agree with the predictions from the undisplaced forces within 0.21% for the chain, bridge and control modes and 1.2% for the contact-N modes; these comparisons test the consistency of the energy derivatives.

**Moving partitions.** Because regional integrals also change when their boundaries move, we report the self-consistent response, $\rho_{\mathrm{scf}}$, and a first-order boundary-corrected response, $\rho_{\mathrm{corr}} = \rho_{\mathrm{scf}} - \rho_{\mathrm{geom}}$, whose last term applies the displaced partitions to the undisplaced density. The correction removes the first-order boundary contribution but not every rigid translation of atomic density; both are sensitivity measures, not physical bounds, and each outcome is stated with its partition and convention.

**Path-polarization criterion.** This predefined criterion requires a path-interior P moment response with the sign of $D_{\mathrm{FD}}$ and magnitude at least $|\eta|$. A path response smaller than half the energy response, $|\rho_M| < |\eta|/2$, fails this criterion; other responses are left unclassified. The criterion supplies no constitutive energy–moment relation, so failing it does not exclude polarization energetics: a hypothetical fixed-coefficient contribution proportional to $-M^2$ changes approximately twice as fast fractionally as $M$ and can remain compatible with the observed chain response.

**Vacancy-interior criterion.** At (1,1), it is met when at least two valid control-to-bridge ratios of charge, spin-magnitude and spectral responses are ≥0.3, and not met when at least two are ≤0.25, provided the partitions agree. Meeting it indicates a spatial response of the vacancy interior, not that this region carries the alignment energy.

**Electronic predictions for the chain.** We predicted that the two-level coupling would decrease by 1–8% (approximately 6% on a kinetic reading) and, as a separately evaluated indicator, that the AP–P Kohn–Sham gap contrast would decrease.

## S6.2. Energy responses of the original three modes

**Table S9. Energy response of the original three modes.** Splittings are in meV per pair, slopes in eV Å$^{-1}$ and $\eta$ is the half-span relative change in $|\Delta E|$. The final column compares the finite-difference and undisplaced-force slopes.

| mode | ΔE(−δ) | ΔE(0) | ΔE(+δ) | η | $D_{FD}$ (eV/Å) | $D_{force}$ (eV/Å) | $|D_{FD} - D_{force}| / |D_{force}|$ |
|---|---|---|---|---|---|---|---|
| (1,1) bridge | +29.41 | +27.64 | +25.75 | −6.61 % | −0.09136 | −0.09139 | 0.023 % |
| (3,0) chain | −57.28 | −50.71 | −44.63 | −12.47 % | +0.3163 | +0.3164 | 0.014 % |
| (1,1) control | +27.12 | +27.64 | +28.14 | +1.85 % | +0.02552 | +0.02557 | 0.18 % |

The energy slopes agree with their force references within 0.2% in all three modes (Table S9). Chain flattening weakens the AP preference; the bridge mode weakens the P preference, and its off-path control strengthens it.

**Chain response.** The half-span change is $\eta = -12.47\%$, and the P-branch couplings decrease by 7.26% and 5.85% in the two spin channels, within the predicted interval, whereas the bond-length-only estimate gives a +8.4% increase. Twice these coupling responses bracket the energy response; this squared-coupling comparison was made after inspection and is not a validated scaling law.

The independent gap-contrast indicator fails: the contrast increases by 5.67% instead of decreasing. After the boundary correction the chain moment decreases by 6.05% (P1) and 5.13% (P2). Both responses lie below half the energy response, 6.24%, in both partitions and also without the correction, so the path-polarization criterion is not met. The P1 margin to this threshold, 0.19 percentage points, is smaller than the conservative whole-cell 3σ of 0.94% but far above the regional 3σ of 0.007%; the P2 margin is 1.10 percentage points against a whole-cell 3σ of 0.77%. This does not show that the chain is energetically passive.

**Bridge and off-path responses.** Tightening the (1,1) bridges gives $\eta = -6.61\%$, while its control gives +1.85%. The bridge-boron moments change by −3.74% (P1) and −2.43% (P2), so the path-polarization criterion is not met; whether the response falls below half the energy response depends on the partition, and the moment response changes sign between the corrected and uncorrected conventions, so no strong causal exclusion follows. The control meets the vacancy-interior criterion, with control-to-bridge ratios of 19.3 (P1) and 7.6 (P2) for vacancy B's charge Q and 0.34 and 0.37 for its spin magnitude |δA|, but these descriptors move in the same direction in both modes while the energy reverses. The spectral item, invalid under its predefined channel-sign condition, is not counted.

The force-difference map (Figure S6c,d) concentrates on the connecting atoms: the five atoms of the bridges or the chain carry 88% and 89% of $\sum_i \left|\mathbf{F}_{\mathrm{AP},i} - \mathbf{F}_{\mathrm{P},i}\right|^2$ for (1,1) and (3,0), a share of force sensitivity rather than of interaction energy. Because this map informed the mode selection, the energy slopes are not blind predictions.

At fixed onsite energies, same-spin occupied–empty mixing lowers the occupied energy, whereas a filled P pair keeps its occupied sum, favouring AP. This is not a theorem for the self-consistent energy: other orbitals, onsite terms, density and double counting respond. The failed gap indicator and the chain-end control's partial empty-channel reproduction leave occupied–empty hybridization a supported candidate, not a measured kinetic-exchange term. Distance or bond length alone fails: flattening shortens the chain bonds yet weakens AP; the closer pair favours P.

## S6.3. Electronic responses of (1,1) to the bridge mode and its control

The bridge and off-path modes change $\Delta E$ by −3.655 and +1.021 meV across the full span. These opposite energy responses test any proposed single-variable descriptor. Table S10 reports the signed spectral contrasts and real-space responses with their numerical floors.

**Table S10. Electronic responses to the (1,1) bridge and off-path modes.** Entries are full-span changes of AP-minus-P contrasts; density entries are percentages, with signed μB values in parentheses. Thresholds combine repeat noise and the rounding of reported values. Classes refer to these two modes only.

| descriptor | bridge | control | threshold | class |
|---|---|---|---|---|
| contact-π splitting contrast (meV) | −7.93 | −10.93 | 0.013 | same direction |
| triad-B e′ exchange-splitting contrast (meV) | +11.64 | +3.69 | 0.95 | same direction |
| triad-A e′ exchange-splitting contrast (meV) | −4.35 | −4.73 | 1.47 | same direction |
| filled-pair contact polarization (μB) | +0.0024 | +0.0018 | 4.6 / 4.4 × $10^{-4}$ | same direction (non-monotonic across −δ, 0, +δ; changes with supercell) |
| contact-N $C_A$ | −1.2 % (+0.0017) | −0.5 % (+0.0007) | floor 2–7 × $10^{-5}$ μB | same direction |
| intermediate-host-shell $C_A$ | +12.7 % (−0.0084) | +2.0 % (−0.0013) | floor 2–7 × $10^{-5}$ μB | same direction |
| contact-N moment contrast, PAW-projected (μB) | +0.002 | 0.000 | 0.003 | rigid (contrast −0.147) |
| π centroid contrast (meV) | +0.757 | −0.362 | 0.13 | moves with ΔE |
| occupied e′ centroid contrast (meV) | +0.17 | −0.78 | 0.063 | moves with ΔE |
| empty e′ centroid contrast (meV) | +1.35 | −1.16 | 0.061 | moves with ΔE |
| $G_{oe}$ (meV) | +1.18 | −0.38 | 0.0055 | moves with ΔE |
| full p×G proxy (meV) | +2.60 | −3.30 | 1.52 | moves with ΔE (through cancellation) |
| even-state tail (share) | +0.0516 | −0.0203 | 0.00084 | moves with ΔE (mainly through the P branch) |
| non-facing-N $C_A$ | +6.7 % (−3.2 × $10^{-4}$) | −11.3 % (+5.4 × $10^{-4}$) | floor 2–7 × $10^{-5}$ μB | \|C\| anticorrelated with \|ΔE\| |
| contact-hollow $C_A$ | +13.7 % (+3.5 × $10^{-4}$) | −1.9 % (−4.8 × $10^{-5}$) | floor 2–7 × $10^{-5}$ μB | \|C\| anticorrelated with \|ΔE\|; control at the cell-scale floor |

In both modes the contact splitting and the spin-magnitude descriptors move in the same direction, or stay below their floors, although the energy response reverses; the PAW-projected contact-moment contrast changes by no more than 0.002 μB. These quantities therefore cannot serve as consistent sole descriptors, although a nearly constant contact contribution to the baseline is not excluded.

The occupied–empty proxies that do reverse are not unique indicators, and the full p×G proxy tracks the energy only through cancellation of components that individually do not reverse; the even-state tail becomes branch-dependent. The non-facing and contact-hollow densities correlate with the preference, but the former reverses under HSE06 and the latter reaches the repeat floor in the control. Individual terms of the Kohn–Sham total-energy decomposition, such as the band-structure energy, are unsuitable indicators: their large changes are compensated by other terms and can depend on the energy reference.

## S6.4. The (3,0) chain and its two controls

The first control tests whether onsite-level changes, rather than coupling along the shortest path, explain the chain response; its rules were fixed before its outputs but after the force slopes were known. It moves B9, B48 and N159, matching the chain in atom count, species and selected bond shortening while lying off every inter-triad path of up to eight bonds; N159 lies on a twelve-bond periodic-image path. The control is thus structurally matched but not electronically identical to the chain perturbation.

The relative onsite shift is the control-to-chain ratio of a selected onsite-level response. $R_{\mathrm{on}}$ uses the relative energy of the chain-coupled and near-degenerate components; $R_c$ uses the chain-coupled component alone. A predefined onsite-leakage correction removes from the coupling ratio the first-order response attributed to the control's own onsite perturbation.

**Table S11. Chain versus off-path response at (3,0).** Ratios compare the control with the chain within each spin channel; ↑ is the occupied and ↓ the empty P frontier channel. The energy response was known from the undisplaced forces and is not a prospective prediction. Coupling ratios are given with and without the onsite-leakage correction.

| quantity | normalization | chain | off-path control |
|---|---|---|---|
| ΔE(−δ, 0, +δ) | meV | −57.28, −50.71, −44.63 | −50.701, −50.713, −50.756 |
| η | half span | −12.47 % | +0.054 % (known beforehand from the forces; not a prospective test) |
| $t_{2L}$, ↑ / ↓ | half span (η scale) | −7.26 / −5.85 % (indicator specified in advance: met) | $R_t$ = −0.031 (↑, specific); ↓ raw +0.275 (intermediate), +0.181 after the predefined onsite-leakage correction (specific) |
| triad Q | control/chain ratio | 1 | 0.018–0.021 (P1/P2, both conventions; specific) |
| triad \|δA\| (robustness) | control/chain ratio | 1 | 0.038–0.040 |
| relative onsite shifts $R_{on}$ (↑ / ↓); chain-coupled component $R_c$ (↑ / ↓) | control/chain ratio | 1 | +0.119 (not met) / −0.297 (met); 0.613 / 0.291 (met) |
| p×G proxy | signed D±, meV (threshold 1.3) | −13.11 | +0.34 |
| $p_{oe}$ | signed D± (full span; η scale) | −0.0040 (−17 %; −8.7 %) | +0.0002, below noise |
| real-space Q, $C_A$, $C_S$ contrasts | full span (η scale) | −11 to −25 % (−5.6 to −12.5 %) | ≤ 1.9 % (≤ 0.95 %) |
| $gap_{AP}$ − $gap_P$ (second indicator specified in advance) | half span (ρ of Section S6.1) | +5.67 % (predicted to fall; failed) | −4.92 % (not part of the control's predefined criteria; descriptive; Figure S7) |

The control meets the stated specificity criterion (Table S11) through the chain-coupled relative onsite shift. Its occupied-channel coupling response is not reproduced (ratio −0.031); the empty-channel raw ratio, 0.275, is intermediate and falls to 0.181 after the predefined correction; the triad density response is approximately 0.02 of the chain's. Specificity thus depends on the correction in one channel. Within first-order single-variable comparisons, the onsite shifts do not account for the chain's energy response; this removes only a restricted alternative, and the failed gap-contrast prediction stands.

**Chain-end control.** The second control addresses chain-end leakage into the coupling descriptor, which the first cannot fully test. It moves B1 and B21, the other boron neighbours of the chain-end nitrogens, without changing chain-interior bond lengths or angles; these atoms lie on six-bond inter-triad paths, so the mode is not strictly off-path. At +0.02 Å it shortens the endpoint bonds by 0.01028 Å (chain: 0.01035 Å) and changes the endpoint angle by −0.708° (chain: −0.701°); its energy-response ratio, approximately 0.0644, was known beforehand from the undisplaced forces.

The onsite rule compares $F = R_E/R_{\mathrm{on}}$ and requires a relative onsite shift of at least 0.25 with $F \leq 0.25$. This factor-four criterion was chosen after the energy ratio was known and is less strict than the factor-ten criterion specified for the first (off-path) control. The coupling rule classifies raw response ratios as specific, intermediate or shared.

All four states meet their acceptance checks. Both P calculations used all-band conjugate-gradient minimization from the outset, because a related reference calculation had failed to converge; we had initially planned to switch the minimizer only for a calculation that had itself failed, and we note this change of plan in the reproducibility record deposited with the raw data.

The full-span energy response is +0.813 meV, compared with +12.652 meV for the chain, giving $R_E = 0.0643$. Its finite-difference slope agrees with the force reference within 0.16%. The relative onsite responses reach 0.333 and 0.578 of the chain values in the two spin channels. The energy-to-onsite ratios are 0.193 and 0.111, meeting the factor-four criterion but not the factor-ten criterion of the first control. The coupling ratio is −0.033 (specific) in the occupied channel and +0.449 (intermediate, with the leakage sign) in the empty channel; triad density responses remain at no more than 0.075 of the chain response.

This control argues against the specified onsite-level shift as the sole first-order explanation of the chain energy response, but nearly half of the empty-channel coupling response can arise from an endpoint perturbation. Onsite leakage and coupling along longer paths both remain compatible; the occupied-channel evidence is the more selective, and the mechanism interpretation remains partly supported. A planned comparison of this control between the CPU and GPU builds of Section S1.1 was not evaluated, and a secondary splitting descriptor was not computed.

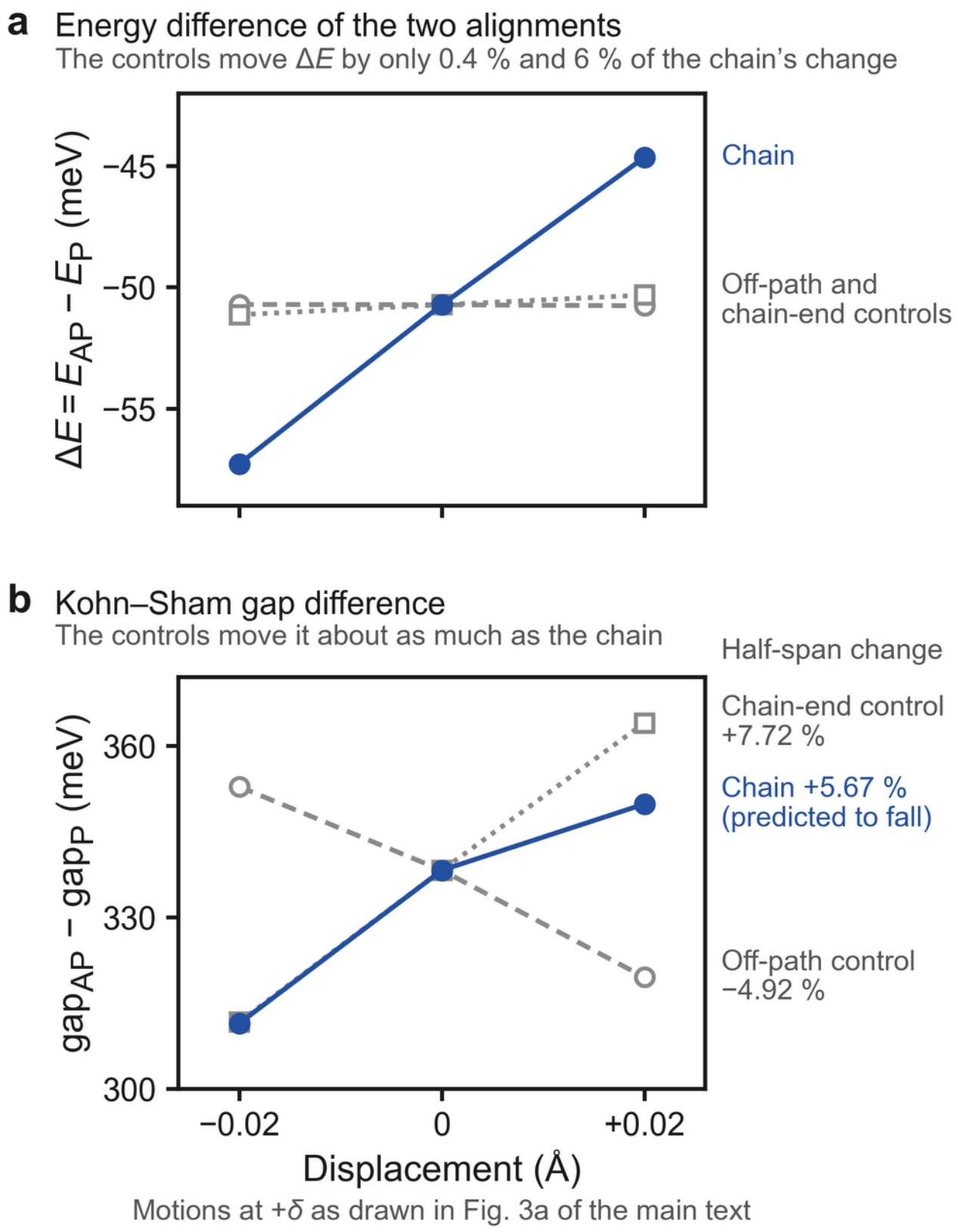

**Figure S7. Energy and gap contrasts under the three (3,0) modes. (a)** The chain changes $\Delta E$ substantially, while the off-path and chain-end controls produce approximately −0.4% and 6.4% of its full-span response. **(b)** The half-span changes of the AP–P gap contrast are comparable for the three modes: +5.67%, −4.92% and +7.72%. The predicted decrease for the chain is not observed; the control values are descriptive only. Source data are deposited (Section S8.3).

## S6.5. Contact-local modes at (1,1)

Because the two original (1,1) modes leave the contact descriptors unable to account consistently for opposite energy responses, we also displaced the contact atoms themselves, with predictions fixed before the eight calculations were performed.

Contact N of A moves N99 towards the midpoint of N100 and N109. Contact N of B moves N100 and N109 towards N99 in a scissor motion. Both shorten the contact, but the force slopes predict opposite effects on $\Delta E$. The modes also change different vacancy interiors and bridge bonds, so the moved side is inseparable from the sign of the energy response in this test.

Three descriptors are specified: the contact $\pi$ splitting contrast $\Delta S_{\pi}$, the local-majority-signed contact moment contrast $\Delta M_C$, and the AP mixing proxy $E_{\text{mix}} = \sum_s p_s G_s$. "Tracks" denotes an above-floor response with the required sign relation to $\Delta E$ in both modes. The contact-polarization hypothesis predicts a negative relation for the splitting and moment contrasts; the mixing hypothesis predicts a negative relation for the mixing proxy and a positive relation for the moment contrast.

All eight states pass the convergence, occupation, total-moment and frontier-pattern checks. Their full-span energy responses are +6.745 meV for contact N of A and −2.490 meV for contact N of B. The energies confirm known force slopes; the descriptor responses are the quantities tested prospectively.

**Table S12. Contact-local descriptor responses.** Entries are full-span changes of AP-minus-P contrasts with their numerical floors. Class labels give the predefined sign relation s and do not identify a separate contact energy.

| descriptor | full-span response, contact N of A / contact N of B | floor | class | predicted by |
|---|---|---|---|---|
| contact π splitting contrast $\Delta S_{\pi}$ (meV) | −24.65 / +4.50 | 0.0132 | tracks (s = −1) | contact polarization |
| contact moment contrast $\Delta M_C$, PAW-projected (μB) | −0.008 / −0.005 | 0.003 | same direction | polarization: tracks (s = −1); hybridization: tracks (s = +1) |
| occupied–empty mixing proxy $E_{\text{mix}}$, AP (meV) | +4.46 / +10.91 | 1.13 | same direction (neither coupling nor onsite-gap part dominates) | hybridization: tracks (s = −1) |

Only the contact $\pi$ splitting tracks (s = −1) in both modes (Table S12); the moment contrast falls and the mixing proxy rises in both. The predefined criterion thus assigns the contact-polarization pattern through the splitting alone, although a side-dependent onsite response can reproduce the same pattern. Because the specification treats the moment descriptor inconsistently, and the same-direction moment change contradicts a purely moment-based polarization account, we adopt the narrower statement that a contact spectral response correlates with the preference without isolating polarization energy.

Across all four (1,1) modes, none of the three descriptors retains one sign relation to $\Delta E$. A nearly constant baseline contribution remains possible, while no evaluated descriptor provides a universal single-variable account of the response. The associated crystal-orbital Hamilton population (COHP) changes (Section S7) are below their numerical floors and do not discriminate between the alternatives.

## S6.6. Transfer of the contact signature to other placements

Predictions for placements not used to define the compact-contact signature were fixed after these data were generated but before the targets were inspected; a whole-cell spin-magnitude quantity had been seen earlier, so the test is temporally separated, not fully blind.

The primary prediction requires $|dA_F| < 0.070\mu_{\text{B}}$ at both other eligible 10×10 placements under both partitions, half the (1,1) contrast; a secondary $\pi$-splitting criterion applies only to a single tracked facing-$\pi$ state. We exclude (2,0), whose branches reorganize, and treat the 12×12 results as descriptive.

**Table S13. Transfer of the facing-atom spin signature.** Contrasts are evaluated at identical nuclei within each placement; P1 and P2 are spatial partitions. The weak (2,2) result is not assigned a preferred alignment.

| placement | supercell | role | ΔE (meV) | shortest facing N–N (Å) | $dA_F$, P1 / P2 (μB) | facing-contrast prediction |
|---|---|---|---|---|---|---|
| (1,1) | 10×10 | calibration | +27.64 | 2.46 | −0.140 / −0.135 | — |
| (3,0) | 10×10 | calibration | −50.71 | 4.89 | −0.052 / −0.050 | (below 0.070) |
| (2,1) | 10×10 | transfer | −7.96 | 4.24 | −0.0097 / −0.0093 | holds |
| (2,2) | 10×10 | transfer | +0.21 (weak, below the 0.5 meV threshold) | 6.54 | +0.00002 / +0.00003 (within the whole-cell repeat floor; 2.4–5.4 times the facing-region floor) | holds |
| (3,1) | 12×12 (different supercell) | reported only | −2.20 | 6.52 | −0.0020 / −0.0019 | holds |
| (4,0) | 12×12 (different supercell) | reported only | −5.24 | 7.40 | −0.0052 / −0.0050 | holds |

Both test placements, (2,1) and (2,2), satisfy the criterion (Table S13): relative to (1,1), the facing contrast is 2.7 times smaller at (3,0) and at least 14 times smaller at (2,1) and (2,2). Because no transferred placement has a single tracked contact-localized filled π state, the secondary criterion is inapplicable, not passed. This supports spatial specificity of the signature, not its energetic origin.

A large top-π splitting alone is not contact-specific (0.196–0.247 eV at other placements versus 0.333 eV at (1,1)); what distinguishes (1,1) is a single localized filled state with a concentrated spin-magnitude contrast. Overlap cancellation also predicts a large contrast at the shortest facing distance, and the small sample cannot distinguish it from contact polarization.

## S6.7. Robustness to geometry, supercell and functional

Functional comparisons use matched coordinates and a PBE reference in the same supercell (Table S14).

**Table S14. Geometry, supercell and functional sensitivity of branch contrasts.** Electronic contrasts are AP minus P at identical coordinates in each column. $R_{\mathrm{AP}}$ denotes a vertical comparison at the AP-relaxed geometry, not the relaxed-branch splitting. HSE06 uses 8×8 for (1,1) and 10×10 for (3,0).

| signature | reference (10×10 PBE, $R_P$) | $R_{AP}$ | 8×8 PBE | HSE06 |
|---|---|---|---|---|
| ΔE (1,1) (meV) | +27.64 | +25.75 | +27.60 | +51.80 (8×8) |
| ΔE (3,0) (meV) | −50.71 | −54.19 | — | −23.36 (10×10, $R_P$) |
| (1,1) contact-π splitting ΔX (eV) | −0.333 | −0.328 | −0.332 | −0.469 |
| (1,1) contact-N $C_A$ (μB) | −0.140 | −0.140 | −0.140 | −0.110 |
| (1,1) contact-N PAW-projected moment ΔX (μB) | −0.151 | −0.144 | −0.147 | −0.120 |
| (1,1) contact-hollow $C_A$ (μB) | +0.0026 | +0.0026 | +0.0025 | +0.0046 |
| (3,0) whole-cell $C_A$ (μB) | −0.127 | −0.133 | — | −0.056 |

The (1,1) contact spin-magnitude signature keeps its sign under all three changes: it persists after AP relaxation, in 8×8 and with HSE06. Relative to matched 8×8 PBE, HSE06 retains 0.74–0.79 of the whole-cell and contact density contrasts and 0.82 of the projected contact-moment contrast, that is, 0.74–0.82 of the matched PBE spin-magnitude and projected-moment contrasts, while the contact π-splitting contrast grows; no single scaling factor applies.

For (3,0), HSE06 opens the same-spin gap from 1.65 to 3.82 eV, reduces the density and projected-spin contrasts to 0.26–0.75 of their PBE values and weakens the AP preference. This is consistent with, but not unique to, a gap-sensitive mixing contribution: similar functional shifts of the two reference splittings admit another description, while the much smaller (2,1) shift rules out a universal constant correction.

Several subsidiary descriptors are not robust: the filled-state contact polarization changes with supercell, the odd/even mixing split is undefined under the HSE06 state tracking, the even-state tail is geometry-sensitive, some (3,0) centroid and gap contrasts reverse, and the (1,1) bridge-B contrast vanishes while its non-facing-N contrast reverses. None of these removes the primary contact signature or the sign contrast.

### S6.8. Limits of the mechanism analysis and further tests

At (1,1), the contact descriptors do not track the four deformation responses consistently, and the displaced side cannot be separated from the sign of the energy response. At (3,0), the predicted gap-contrast decrease fails, and the chain-end control reproduces part of the empty-channel coupling. All predefined COHP groups lie below their numerical floors, the (3,0) B–N chain sensitivity check fails and the contact-window signs are not uniformly consistent, so below-floor contrasts neither establish absent bonding contributions nor support contact polarization by elimination.

The path-moment rule specifies no energy–moment relation, and polarization and hybridization can co-vary. Regional descriptors depend on the partition, PAW-projected moments are not quantum spins and the occupied–empty argument holds only in a fixed-level model. We use one charge state and principally collinear PBE; the energy slopes were known from the undisplaced forces beforehand, several descriptor refinements are retrospective and individual terms of the total-energy decomposition depend on the energy reference. None of these limits changes the calculated branch ordering.

Hybrid-functional deformations, a second supercell, further compact placements, more selective structural or electronic constraints, collective vacancy-spin constraints and a completed larger-cell angular series, none of which we have performed, would separate onsite, hybridization and polarization responses.

At (3,0), chain flattening weakens the AP preference and the same-spin coupling and neither control reproduces the occupied-channel response, partly supporting an occupied–empty hybridization contribution, qualified by the failed gap indicator, composite modes and co-varying polarization; the exploratory chain N–N bonding contrast (Section S7) is not a share of $\Delta E$.

At (1,1), the contact signature is robust and spatially specific, but its splitting reports a spin-dependent potential, not an energy gain, and contact polarization, bonding redistribution and overlap cancellation remain compatible candidates.

## S7. Bond-resolved analysis (exploratory)

Crystal-orbital Hamilton populations (COHP) and their integrals (ICOHP) compare occupied one-electron bonding in P and AP. They do not decompose the branch energy difference, the exchange coupling or the exchange–correlation energy, and in a vacuum slab a well-reproduced occupied state does not guarantee a well-conditioned local Hamiltonian. No matched P/AP comparison of crystal-orbital overlap populations (COOP) or bond indices (COBI) is available.

We evaluate them with LOBSTER 5.1.1.[15,16] Every bond group used in a predefined test lies below its numerical floor. One exploratory chain N–N contrast exceeds its threshold, but it does not establish sensitivity of the predefined chain B–N test or identify a magnetic energy contribution.

Because the predefined rules required a controlled bonding response above its floor, neither a contact-bonding interpretation nor the absence of a bonding contribution is established, and below-floor bonding changes do not support contact polarization by elimination.

**Method.** For this analysis the wavefunctions were recomputed with 848 bands, more than the 792 B/N 2s2p local basis functions per spin, and reproduce the reference energies. The primary local basis is the PBE-fitted LOBSTER basis set, with the Bunge basis as an alternative, and each spin channel is integrated up to the midpoint of its own gap. An error in our extraction of the COHP data was corrected before the affected values were evaluated; the test criteria were not changed.

**Baseline contrasts.** Contrasts are P minus AP in spin-summed occupied bonding. The predefined chain B–N sensitivity check fails: its −0.0694 eV contrast is about one tenth of the 0.6695 eV threshold (three times the floor), so the method cannot detect that group at the required sensitivity.

The contact N–N contrast is below threshold and its defect-window contribution has the opposite sign at all tested window edges; the chain B–N bonds have mixed signs.

The two bridged N–N pairs of the (3,0) chain give an exploratory contrast of −0.0403 eV against a 0.0145 eV threshold; we selected this set after inspecting the results, and it enters none of the predefined tests.

**Contact modes.** In both contact modes every bond response stays below its floor; the contact-group floor is 76–86% of the group's entire ICOHP, so sensitivity to a small contact contrast is poor. Our predefined joint criterion would assign the contact-polarization label whenever all bonding responses lie below their floors; we do not adopt that label, because below-floor responses cannot exclude a bonding contribution.

**Projection quality.** We reconstruct the local Hamiltonian from a 792-band window. In the 25 Å vacuum slab, 196–209 empty bands carry less than 5% in-sphere weight, and empty-block overlap deviations of 0.05–2.50 break the σ completeness sum rule. This ill-conditioned Hamiltonian enters the occupied ICOHP values and sets the large numerical floors for bonds far from the defects; total spilling is approximately 37%, against charge spilling of approximately 1.12–1.13% per spin. A residual of approximately 0.011 eV per bond per spin, common to both bases, is assigned neither solely to the reconstruction nor solely to the occupied coefficients; agreement between the two bases is therefore not independent evidence. Two diagnostic calculations distinguish these sources only partially, and we do not claim to have removed the problem. Four noise models introduced after inspection change none of the predefined classifications; only a contact-side N–N set outside the predefined tests lies above some of their thresholds.

The (1,1) deformation states recomputed with the CPU build of Section S1.1 have larger empty-block overlap deviations and larger bond-analysis floors than those computed with the GPU build; projecting the same wavefunctions in the computing environment used to project the GPU-build states, which we have not done, would be needed to isolate the source of this dependence.

**Deformation groups.** Every predefined bridge, chain and off-path deformation group stays below its original and retrospective floors; no bond analysis exists for the chain-end control.

**Chain N–N decomposition (exploratory).** Within the chain N–N contrast, N99–N109 and N109–N119 contribute −0.0206 and −0.0197 eV, mainly from occupied defect-derived levels, with a threshold-dependent decomposition. Chain flattening changes it by approximately +0.0100 eV, exceeding only the retrospective N–N noise and not the original floor; the off-path change is approximately −0.0021 eV. It is a projected occupied-bonding contrast, not a magnetic energy or a unique mechanism.

**Transfer placements.** Transfer COHP at (2,1) and (2,2) is descriptive, since no quantitative prediction was defined, and every set is below threshold.

# S8. Reduced models, analysis chronology and data

## S8.1. Reduced models (exploratory)

These exploratory models are kept separate from the accepted DFT energy map. Wannier90 3.1.0 constructs six dangling-bond functions per spin channel from nitrogen $sp^2$ trial orbitals, with an outer window from $E_F - 5$ to $E_F + 3$ eV and no inner window, following Marzari et al. and Pizzi et al.[17,18]

Local blocks separate approximately into $a_1{}'$ and $e'$ sectors. The AP $e'$-to-$e'$ hopping norms are 140, 101, 217 and 5 meV for (1,1), (2,1), (3,0) and (2,2). Their ordering is informative, but the model is not converged: fourteen of sixteen spin-channel disentanglements reach the iteration limit, and deviations from the Kohn–Sham spectrum are 101–166 meV.

Window and localization variants change the hopping ratios by approximately 15%, and inner-window or guiding-centre changes by about a factor of two, so we use the hoppings only as qualitative descriptors. A restricted model with $J_{\mathrm{kin}} = -\| t_{ee} \|^2 / U_{\mathrm{eff}}$ and positive denominator necessarily gives non-positive $J$ and cannot explain the P preference by itself. Calibrating the denominator at (3,0) or at (2,1) leaves (1,1) residuals of approximately +24.4 and +21.5 meV. The ±40% bands of this comparison are illustrative, not confidence intervals, and the residuals are model mismatches, not measurements of direct exchange.

A six-orbital, eight-electron model with constrained-RPA interactions and exact diagonalization is conditional on its one-electron reference. Its response cut-off series is not converged (1.122191 meV against a 0.10 meV target), and we rejected the (2,1) constrained-RPA tensors because transitions within the target manifold dominate their screening. No physical mapping is accepted, and the reference was not chosen to match the DFT sign.

Changing only the one-electron reference at the fixed (1,1) tensor changes the exact-diagonalization parameter $J_{21}$ from +16.750455 meV under partial-Hartree subtraction to −316.285222 meV under model-HF subtraction; neither prescription is independently established as the physical environmental contribution.

A quantitative material model would require a defensible projected exchange–correlation or environmental construction, including PAW augmentation.

## S8.2. Analysis chronology

We distinguish tests specified before generating their outputs from analyses developed after examining existing states. The deformation modes, electronic acceptance conditions, selected descriptor predictions, angular hold-out rules and constraint-radius test were documented in advance of their respective calculations. Table S15 and the reproducibility record deposited with the raw data (Section S8.3) document this chronology, including amendments and uncompleted tests, and record which rules were specified before their outputs existed and which were developed after inspection.

**Table S15. Timing of definitions and predictions.** "Specified in advance" refers to the outputs tested, not to all earlier exploratory information. Qualifications identify energy slopes known from forces, retrospective thresholds and later amendments.

| analysis | specified before its outputs existed? | qualifications | Section |
|---|---|---|---|
| (1,1) bridge, (3,0) chain and (1,1) off-path deformation modes: acceptance conditions and descriptor rules | yes | the energy slopes were known from the undisplaced forces beforehand, and the modes were chosen after the force map had been seen | S6.1–S6.2 |
| (3,0) off-path control | yes | The energy response was known from the forces; specificity in the ↓ (empty) channel depends on the predefined onsite-leakage correction | S6.4 |
| (3,0) chain-end control | yes | energy ratio known beforehand; its factor-four criterion was specified after that ratio was known | S6.4 |
| (1,1) contact modes (contact N of A, contact N of B) | yes | energy responses known from the forces; no control; the displaced side cannot be separated from the sign of the energy response | S6.5 |
| transfer test at (2,1) and (2,2) | Predictions were fixed before inspection of the target descriptors, but after the outputs had been generated | A whole-cell spin-magnitude quantity was inspected before the predictions were fixed; the target descriptors were examined later | S6.6 |
| spectral analysis of the existing states | no | written after the states had been observed; three amendments followed data, the third explicitly post hoc | S4.4, S6.3 |
| angular hold-out, 112.5° rule, constraint-radius rule, (1,1) 10×10 rule | yes | the 112.5° point was accepted after the hold-out failed; the global-rotation criterion at 90° was written after the 0° result; one auxiliary acceptance criterion was loosened after inspection | S5.2, S5.4 |
| bond groups and numerical floors of the bond analysis | yes | corrections to the COHP data extraction made before the affected values were evaluated; the corrected noise models and the chain N–N set are post hoc | S7 |
| operational 5 % moment criterion (Table S2) | no | written after the moments were known | S2.1 |

For the deformation rules, the undisplaced forces had already revealed the first-order energy slopes used to choose the modes. The transfer-test predictions were fixed after the numerical outputs had been generated but before the target descriptors were examined; the test is therefore temporally separated from its targets but not fully blind, because a whole-cell spin-magnitude quantity had already been inspected. The spectral analysis of the existing states, the operational 5% moment criterion and the factor-four energy-to-onsite criterion of the chain-end control (Section S6.4) are exploratory or retrospectively calibrated. The corrections to the COHP data extraction preceded the values they affect, whereas the noise models, the chain N–N set and the odd-channel mixing analysis are post hoc. We claim no complete preregistration, and where a recorded outcome label is broader than the data justify, especially below the numerical floor, we adopt the narrower physical statement. Planned but not completed were the strength-20 endpoint check and second path (Section S5.2), the 10×10 (1,1) AP endpoint (Section S5.4), a comparison of the chain-end control between the CPU and GPU builds (Section S6.4) and three bond-analysis checks: the chain-end control, a matched P/AP COOP comparison and a bond analysis of the CPU-build wavefunctions in the computing environment used to analyse the GPU-build states (Section S7).

## S8.3. Data availability

The source data underlying the figures and tables, the raw data and a separate reproducibility record are deposited under DOI 10.5281/zenodo.22994971. The source data identify the plotted or tabulated values, units and originating calculations; the reproducibility record lists calculation records, input checksums and the test chronology. The deposition is not a complete archive of wavefunctions and charge densities.

Scripts are available from the author on reasonable request, and the licensed VASP PAW potentials are not redistributed.